# SOLID-LIQUID EQUILIBRIA FOR THE BINARY SYSTEMS NAPHTHALENE OR BIPHENYL + 1-TETRADECANOL OR + 1-HEXADECANOL

Luis Felipe Sanz, Juan Antonio González,* Fernando Hevia, Isaías. García de la Fuente, and José Carlos Cobos

*corresponding author, e-mail: jagl@termo.uva.es; Fax: +34-983-423136; Tel: +34-983-423757

G.E.T.E.F., Departamento de Física Aplicada, Facultad de Ciencias, Universidad de Valladolid, Paseo de Belén, 7, 47011 Valladolid, Spain

**Abstract**

A differential scanning calorimetric technique has been used to obtain solid-liquid equilibrium temperatures for the mixtures naphthalene or biphenyl + 1-tetradecanol, or + 1-hexadecanol. All the systems show a simple eutectic point, whose final composition was determined by means of the Tamman's plots using the needed values of the eutectic heat and of the heat of melting, which are also reported. DISQUAC interaction parameters for the OH/aromatic contacts in the selected systems are given. The present experimental SLE phase diagrams are similarly described by DISQUAC and UNIFAC (Dortmund) models. However, the comparison of DISQUAC and UNIFAC results for systems involving naphthalene and shorter 1-alkanols (methanol-1-octanol) reveals that the temperature dependence of the interaction parameters is more suitable in DISQUAC. The systems are also investigated in terms of the concentration-concentration structure factor. It is shown that the positive deviations from the Raoult's law of the studied solutions become weaker when the homocoordination decreases.

**Keyword/s**

SLE/ DSC/PAH 1-alkanols/group contributions

## 1. Introduction

Solid-liquid equilibria (SLE) measurements are of great relevance for industrial design and development based on melting crystallization [1,2,3,4]. SLE data have also gained interest due to their applications in the field of the accumulation and transference of thermal energy. Such applications are based on the heat absorbed and released during a physical state change (typically, solid-liquid transitions). Thus, phase change materials (PCMs) are very efficient heat accumulators since they show the mentioned transitions with large latent heats and a great ability to stabilize the temperature in a narrow range [5,6]. The latter can be attained by means of the creation of mixtures at the eutectic composition. The PCM enclosed in a matrix and suspended in a liquid material can also serve as a high-performance heat-transfer fluid with heat capacity increased by the latent heat of PCM. Another application of the SLE data is the investigation and determination of suitable deep eutectic solutions (DESs) [7,8], which are mixtures of two or three components composed of a certain proportion of hydrogen bond acceptors and hydrogen bond donors. DESs may replace successfully ionic liquids (ILs), due to they have low vapor pressure, stable chemical properties, low melting point and no combustion support. In addition, compared with ILs, DESs have other advantages such as simple synthesis, low toxicity and low price. They can be used, for example, for the extraction of metal ions from spent batteries [9,10].

From the experimental point of view, in this work we have determined by differential scanning calorimetry (DSC) the SLE curves of the eutectic mixtures naphthalene, or biphenyl + 1-tetradecanol, or + 1-hexadecanol. SLE data, obtained by a synthetic method, for naphthalene + 1-alkanol mixtures [11,12] and for biphenyl + 1-octadecanol, or + 1-icosanol systems [13], determined by DSC, are available in the literature. SLE data for solutions containing alkanols with a large number of C atoms are relevant in fat, cosmetic and oil technology. This type of 1-alkanols have also been studied as PCMs [14]. On the other hand, biphenyl and naphthalene are both polycyclic aromatic hydrocarbons (PAHs) built by blocks of benzene. The investigation of solutions with these compounds is needed for a better understanding of aromatic-aromatic interactions, commonly encountered in very complex systems [15]. Due to its stability and inertness, biphenyl is employed as heat-storage material [16], and the eutectic mixture diphenyl ether + biphenyl is used as heat transfer agent [17]. In addition, PAH molecules can be found in many materials, such as coal heavy petroleum products or lubricants, and in many processes related to chemical transformations of crude oil or incomplete combustion of hydrocarbon fuels. The study of PAH mixtures is also useful to achieve a suitable description of heavy petroleum fractions, needed to avoid flocculation and deposition of asphaltenes [18,19], a crucial problem during the exploitation, transport and storage of crude oil.

The current SLE measurements have been correlated using the expressions for activity coefficients derived within the Wilson [20] and NRTL [21] models. In order to investigate the ability of group contribution methods to describe SLE phase diagrams of this type of systems, the UNIFAC [22] and DISQUAC [23,24] models have also been applied, and their results analysed by considering the different combinatorial term and temperature dependence of the interaction parameters of each model. Along UNIFAC calculations, the mentioned parameters were taken from the literature [25]. Two studies using DISQUAC on naphthalene + 1-alkanol (up to 1-octanol) mixtures [26] and on biphenyl + 1-octadecanol or + icosanol systems [13] have been previously provided. Such studies were the continuation of an initial work on the characterization in terms of DISQUAC of binary systems formed by benzene and long chain 1-alkanols using SLE data [27]. The present measurements allow a more accurate determination of the DISQUAC interaction parameters for OH/aromatic contacts in the selected systems. On the other hand, they also facilitate to investigate how deviations from the Raoult's law change when the 1-alkanol size increases in mixtures with a given aromatic compound, or when, in solutions with a given 1-alkanol, the aromatic surface increases, i.e., when, e.g., benzene is replaced by naphthalene. These matters are investigated not only on the basis of the SLE data, but also through the application of the formalism of the concentration-concentration structure factor, $S_{CC}(0)$ [28,29], a method concerned with the study of fluctuations in the number of molecules in a given binary mixture regardless of the components, the fluctuations in the mole fraction and the cross fluctuations. In this work, $S_{CC}(0)$ is determined using DISQUAC.

## 2. Experimental

### *2.1 Materials*

All the compounds were used as received, without further purification. Table 1 shows their source and purity, determined by gas-liquid chromatography by the manufacturer.

**TABLE 1 goes here**

Table 2 lists their physical properties measured in this work: $T_m$, melting temperature and $\Delta H_m$ molar enthalpy of fusion. Our values are in good agreement with experimental results from the literature. Values of the change of the molar heat capacity during the melting process, $\Delta C_{pm}$, for biphenyl, 1-tetradecanol and 1-hexadecanol are also included in Table 2.

**TABLE 2 goes here**

### *2.2 Apparatus and procedure*

The samples were prepared using an analytical balance Sartorius NSU125p (weighing accuracy $\pm 10^{-8}$ kg). Mole fractions were calculated on the basis of the Relative Atomic Mass Table of 2015 issued by the Commission on Isotopic Abundances and Atomic Weights (IUPAC)] [30]. The mixing was carried out by placing the system components in a closed glass flask, and heated them, under continuous stirring, up to 20 K above the highest melting temperature of the involved compounds so that a homogeneous liquid mixture was obtained. The system was hold at that temperature during at least 10 minutes. This method is useful to eliminate thermal memory. Then, the sample was rapidly cooled to crystallization and a small amount, about 10-15 mg, was inserted and hermetically crimped into a Tzero aluminium pan. The magnitudes $T_{\mathrm{m}}$ and $\Delta H_{\mathrm{m}}$ were determined using a DSC TA Instrument, model Q2000, equipped with a refrigerated cooling system RSC90 which allows operations from $T$ = (183.15 to 673.15) K. All the measurements were performed under a nitrogen atmosphere with a volumetric flow rate of 50 $\mathrm{cm^3\ min^{-1}}$ at a constant heating rate of 0.5 $\mathrm{K\ min^{-1}}$. The equipment was calibrated following the procedures indicated by the manufacturer using 99.99% pure indium ($T_{\mathrm{m}}$ = 429.7485 K; $\Delta H_{\mathrm{m}}$ = 3.281 $\mathrm{kJ\ mol^{-1}}$ [31]). For the systems under consideration, at a given composition, the corresponding thermograms show two peaks (Figure S1, supplementary material). One of them is more or less independent of the molar fraction and corresponds to the eutectic temperature. The other one occurs at the solid-liquid equilibrium temperature. The molar enthalpies of the transitions (fusion/eutectic), of the pure compounds and of the mixtures, can be determined as the area limited by the heat flow curve and the baseline built between the points where this transition occurs. It is to be noted that transition temperatures ($T_{\mathrm{tr}}$), close to the melting point, have been reported for 1-tetradecanol ($T_{\mathrm{tr}}$ = 311 K [32], $T_{\mathrm{m}}$ = 310.75 K (this work)) and for 1-hexadecanol ($T_{\mathrm{tr}}$ = 321.1 K [33] or 322.1 [34]; $T_{\mathrm{m}}$ = 321.8 K (this work)). In the present case, these transitions are overlapped with the melting process as it is indicated by the peaks of pure 1-alkanols, which are wider than those of biphenyl or naphthalene (Figure S1).

The estimated uncertainties for mole fraction and temperature measurements are 0.0005 and 0.3 K, respectively. The relative uncertainty of the enthalpy values is estimated to be 0.03.

### *2.3 Experimental results*

Table 3 contains our experimental solid-liquid equilibria temperatures, $T_{\mathrm{SLE}}$ vs. the mole fraction of the involved PAH (see Figures 1,2,3,4). All the systems show a simple eutectic point. No data have been encountered in the literature for comparison. Nevertheless, it must be mentioned that both the composition and temperature of the eutectic points of biphenyl + 1-alkanol mixtures increase with the alkanol size (Figure S2). For example, in the case of the eutectic temperature, we have: 307.2 K (1-tetradecanol), 316.8 (1-hexadecanol) (this work); 321.5 [13], 323.0 [35] (1-octadecanol), 328.7 (1-icosanol [13]). The equation of the solid-equilibrium curve of a pure solid component i [36]:

$$-\ln x_{\mathrm{i}} = (\Delta H_{m\mathrm{i}} / R)\left[1/T - 1/T_{m\mathrm{i}}\right] - (\Delta C_{Pm\mathrm{i}} / R)\left[\ln(T/T_{m\mathrm{i}}) + (T_{mi}/T) - 1\right] + \ln \gamma_{\mathrm{i}} \quad (1)$$

**TABLE 3 goes here**

In this equation, $x_i$ is the mole fraction and $\gamma_i$ the activity coefficient of the component i in the solvent mixture, at temperature $T$. The physical constants, needed for calculations, are listed in

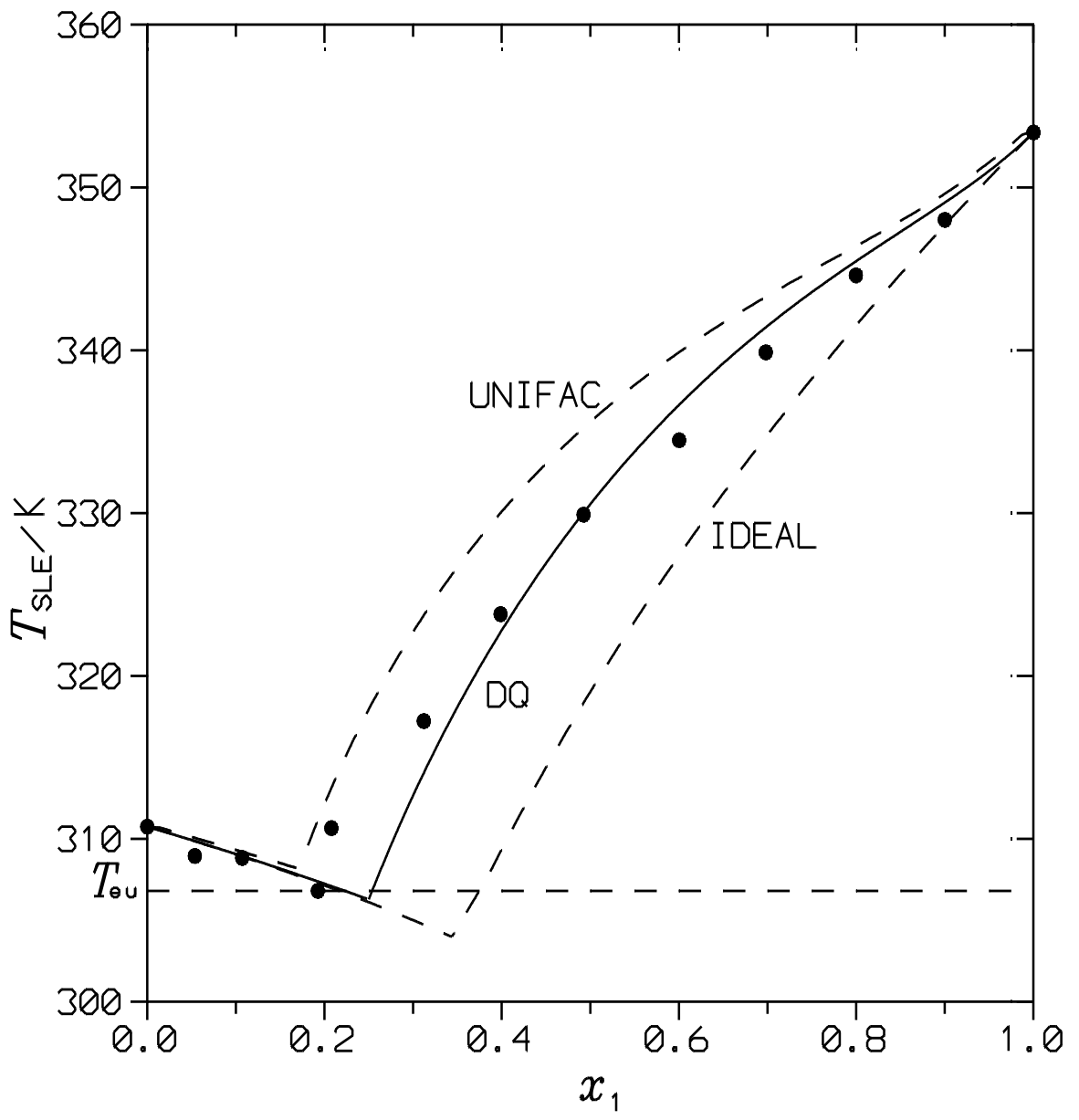

**Figure 1** SLE phase diagram for the naphthalene(1) + 1-tetradecanol(2) mixture. Points, experimental results (this work). Lines, results from the application of different models: DISQUAC (DQ), UNIFAC and Ideal Solubility Model (IDEAL). The eutectic temperature, $T_{eu}$, is also indicated.

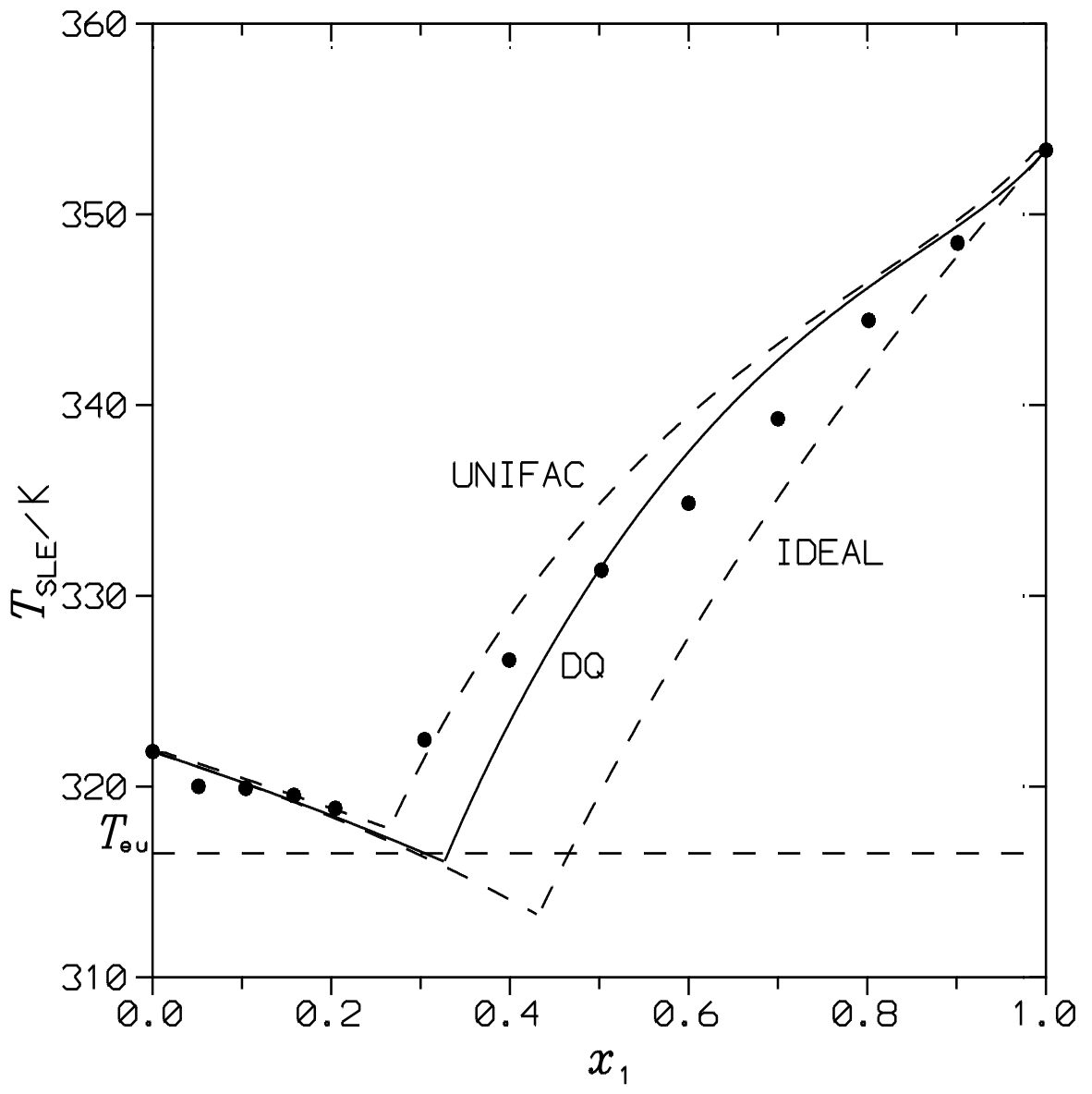

**Figure 2** SLE phase diagram for the naphthalene(1) + 1-hexadecanol(2) mixture. Points, experimental results (this work). Lines, results from the application of different models: DISQUAC (DQ), UNIFAC and Ideal Solubility Model (IDEAL). The eutectic temperature, $T_{eu}$, is also indicated.

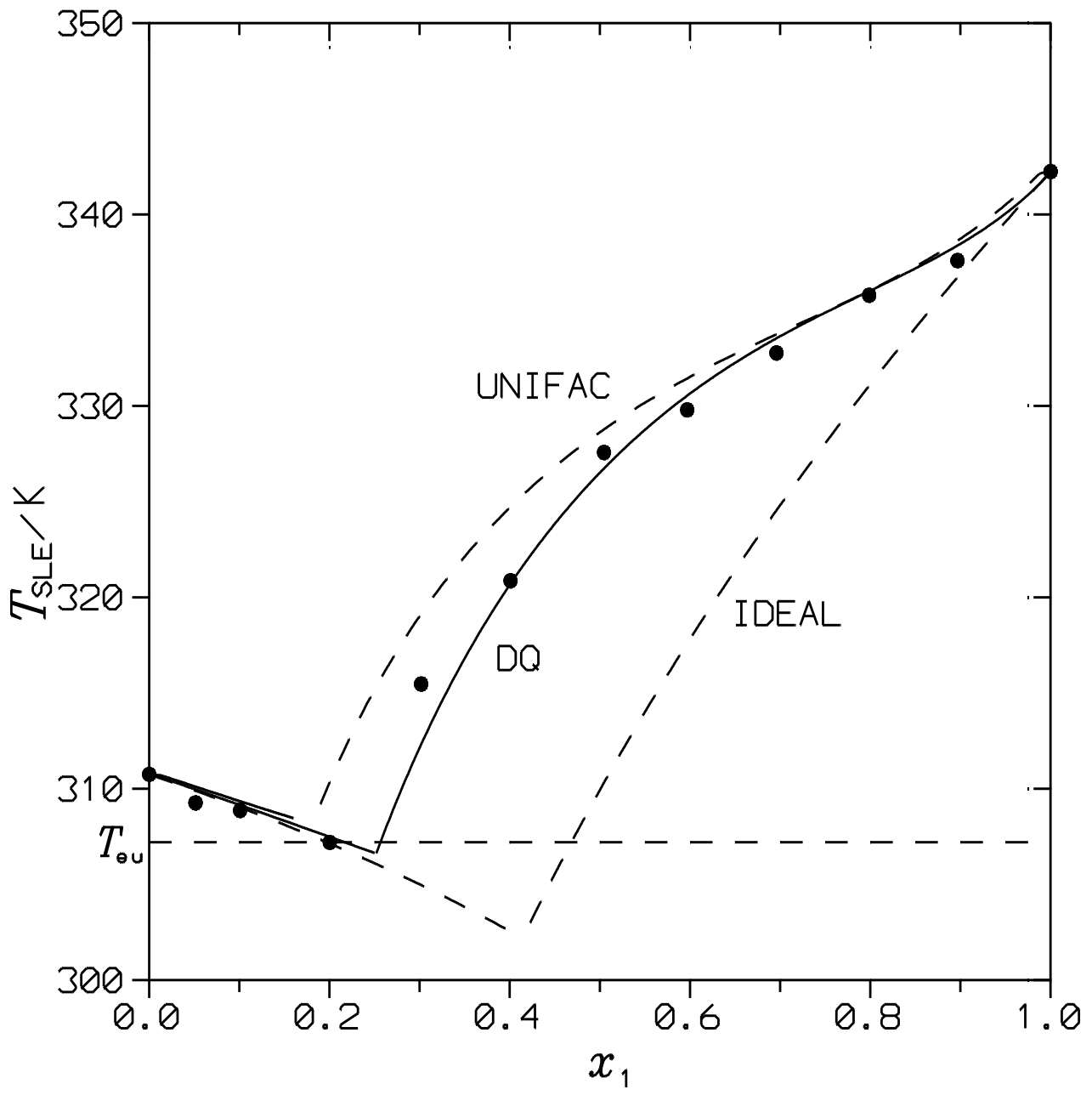

**Figure 3** SLE phase diagram for the biphenyl(1) + 1-tetradecanol(2) mixture. Points, experimental results (this work). Lines, results from the application of different models: DISQUAC (DQ), UNIFAC, and Ideal Solubility Model (IDEAL). The eutectic temperature, $T_{eu}$, is also indicated

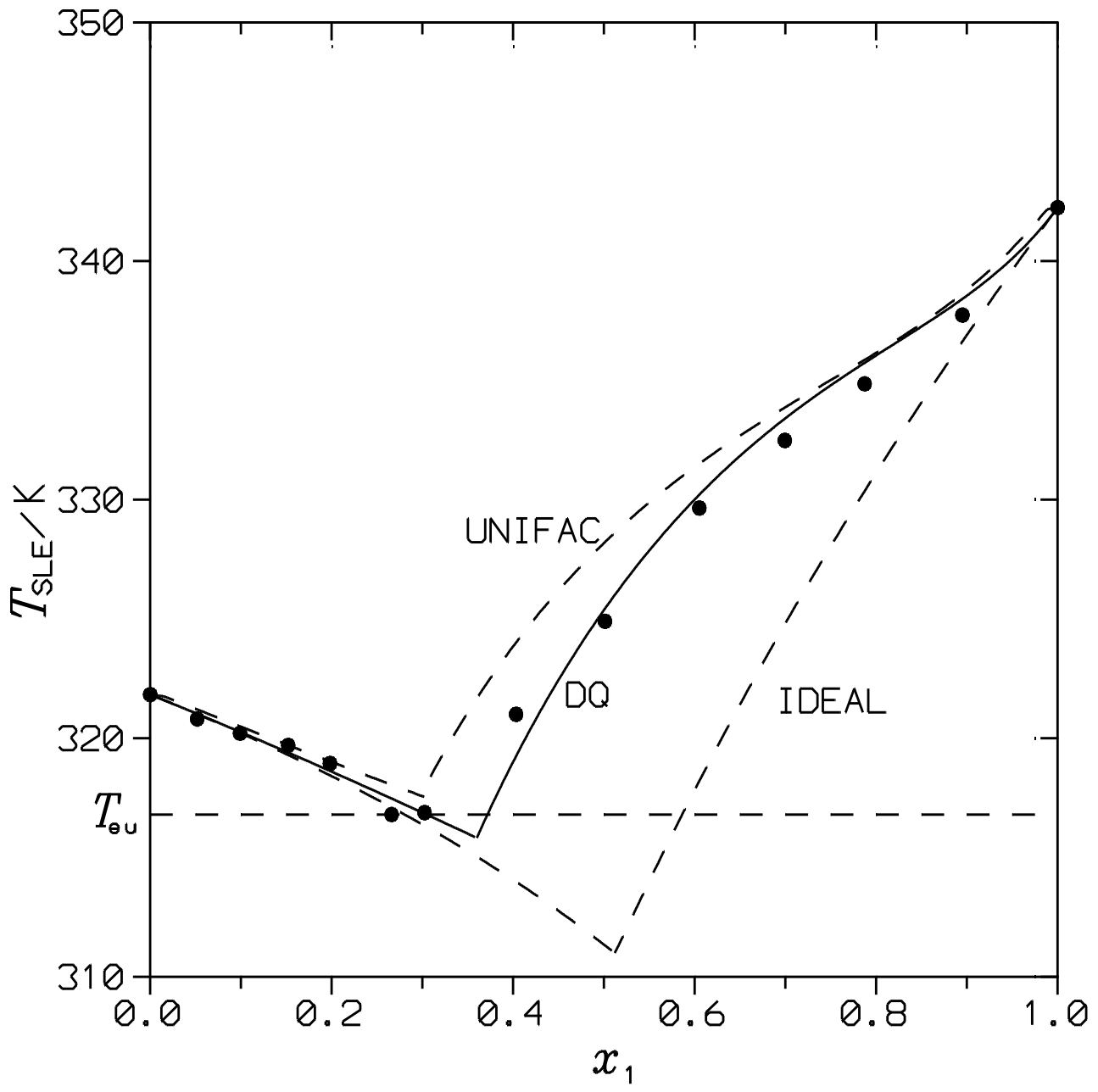

**Figure 4** SLE phase diagram for the biphenyl(1) + 1-hexadecanol(2) mixture. Points, experimental results (this work). Lines, results from the application of different models: DISQUAC (DQ), UNIFAC, and Ideal Solubility Model (IDEAL). The eutectic temperature, $T_{eu}$, is also indicated

Table 2. Values of $\gamma_1$ obtained from equation (1) are shown in Table 3 (see also Figure S3). The eutectic heats ($\Delta H_{eu}$) and the heats of melting ($\Delta H_m$) are collected in Table 4. These values

**TABLE 4 goes here**

have been used to determine the final composition of the eutectic points (Table 3) on the basis of the Tamman's plots [37,38,39] (Figures 5-6). In view of these results, it seems that the present eutectic mixtures may be potential PCMs.

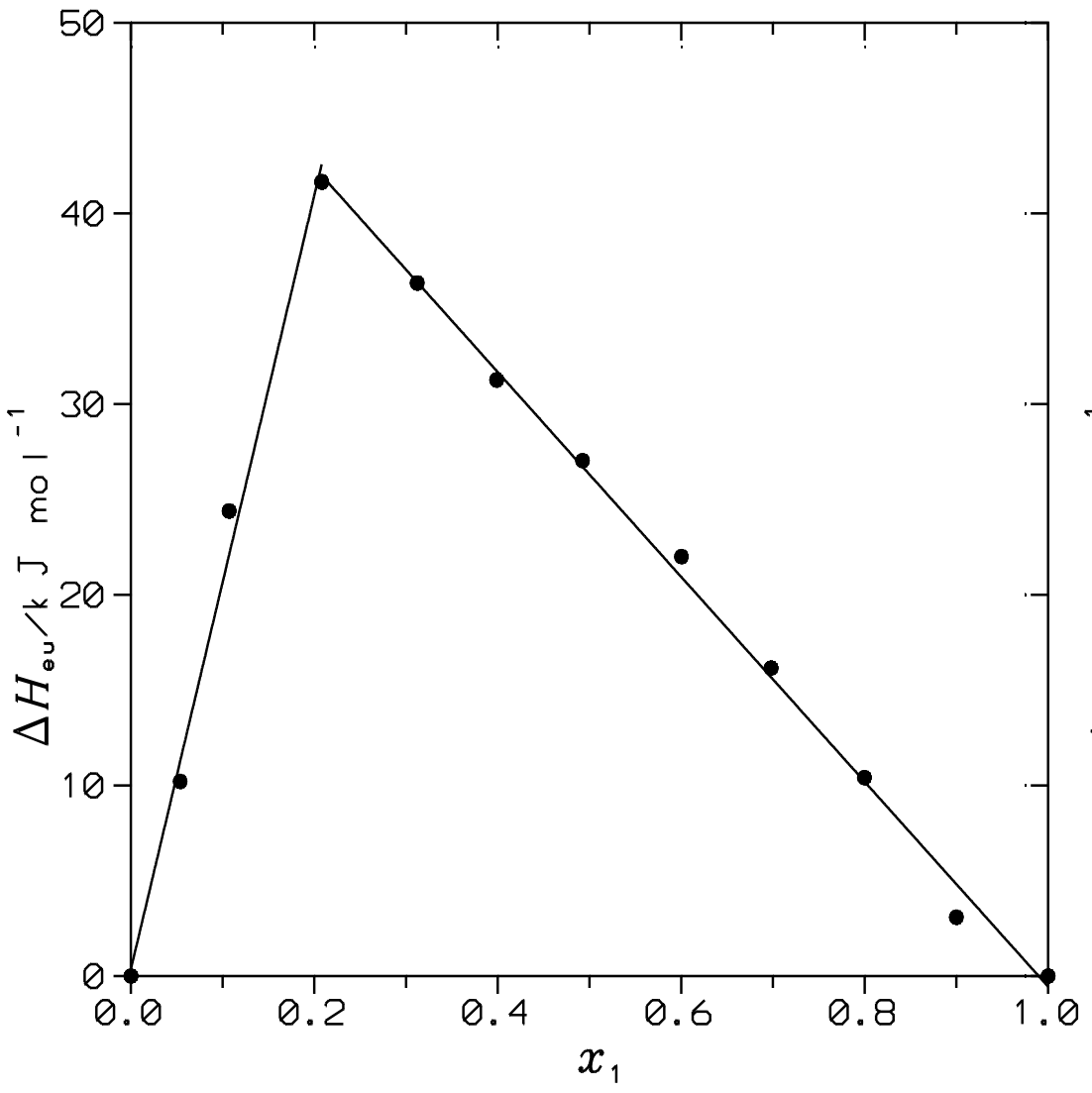

Figure 5a

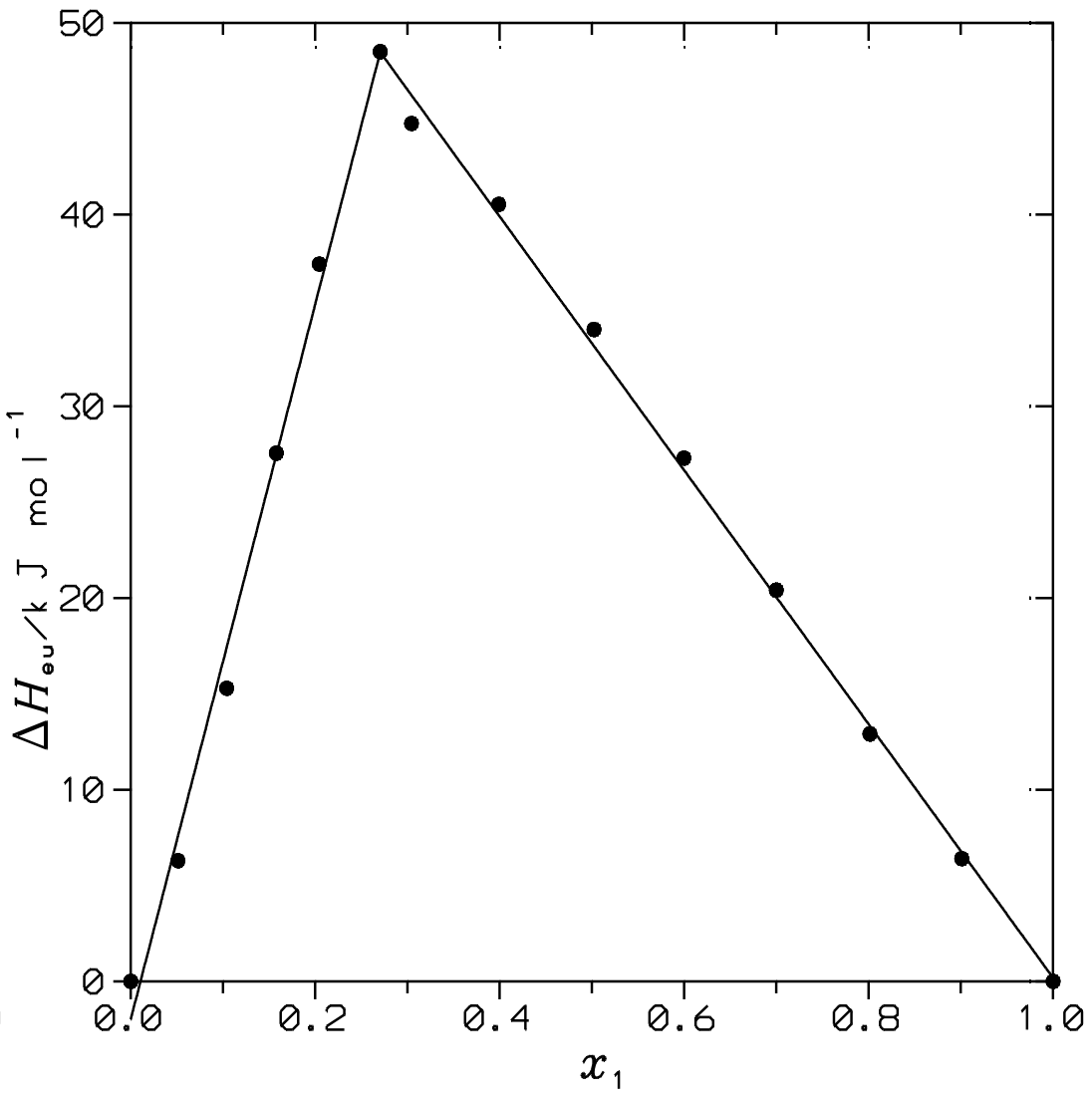

Figure 5b

**Figure 5** Tamman's plots for the mixtures naphthalene(1) + 1-tetradecanol(2) (Figure 5a), or + 1-hexadecanol(2) (Figure 5b)

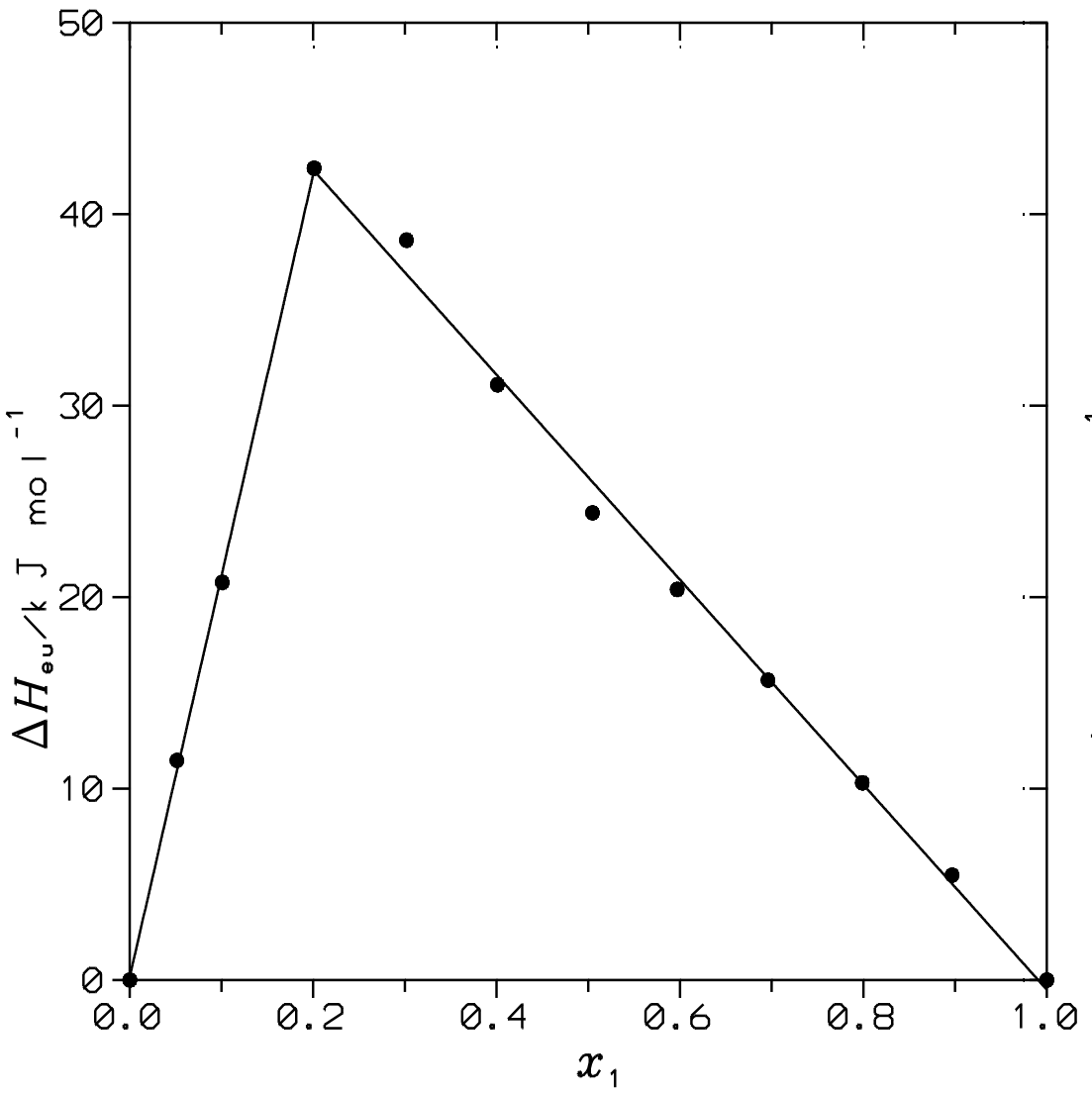

**Figure 6a**

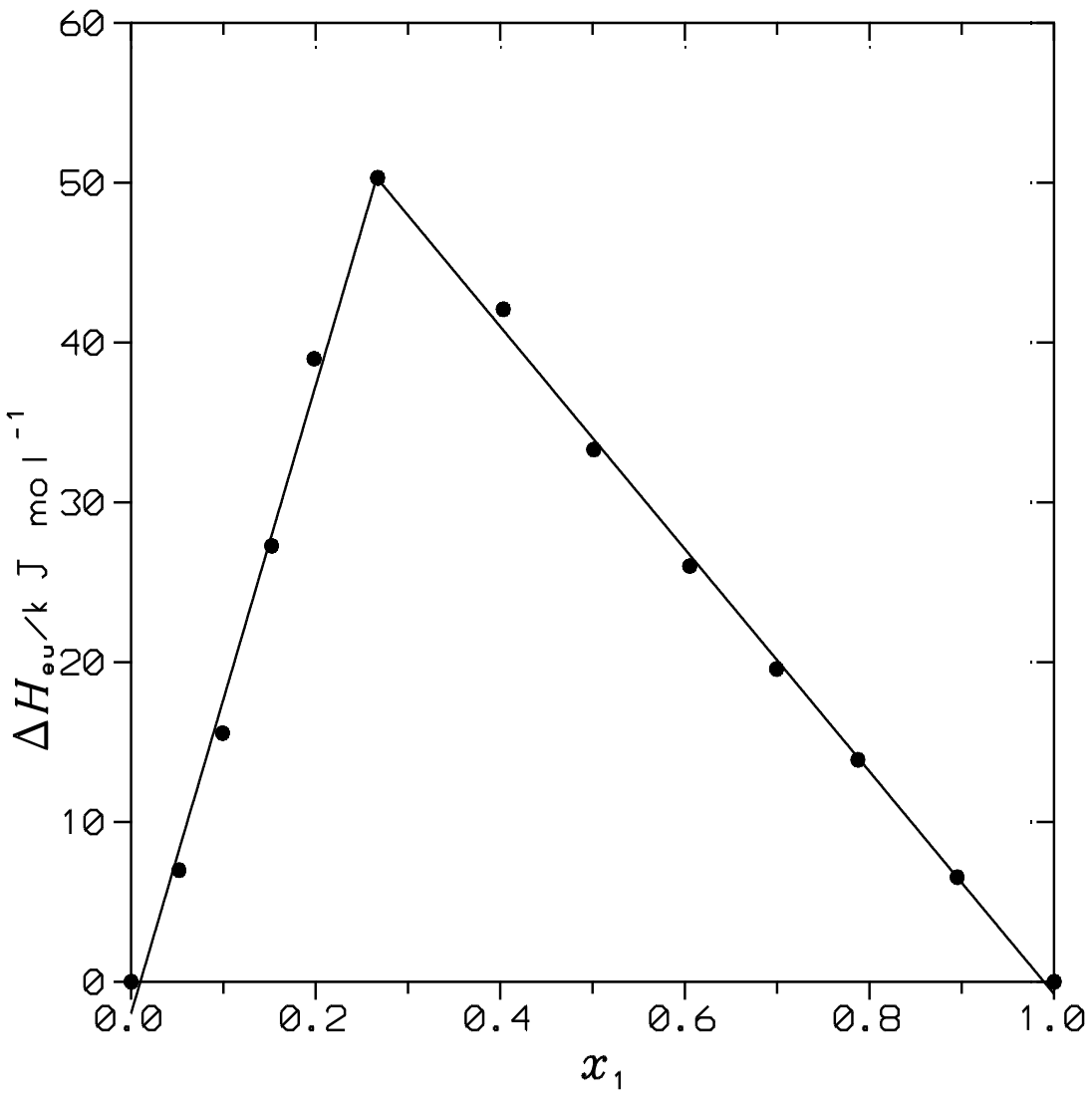

**Figure 6b**

**Figure 6** Tamman's plots for the mixtures biphenyl(1) + 1-tetradecanol(2) (Figure 6a), or + 1-hexadecanol(2) (Figure 6b)

## 3. Models

### *3.1 Correlation equations*

The Wilson equation [20] and the NRTL model [21], both based on the local composition concept, have been applied to correlate the ($x_1$,$T$) data. For a binary mixture, the Wilson equation for the molar excess Gibbs energy, $G_\mathrm{m}^\mathrm{E}$, is:

$$G_\mathrm{m}^\mathrm{E}/RT = -x_1\ln(x_1 + x_2\Lambda_{12}) - x_2\ln(x_2 + x_1\Lambda_{21}) \qquad (2)$$

where $\Lambda_\mathrm{ij} = \frac{V_\mathrm{mj}}{V_\mathrm{mi}}\exp(-\frac{\lambda_\mathrm{ij}-\lambda_\mathrm{ii}}{RT})$ (i,j =1,2), being $V_\mathrm{mi}$ the molar volume of component i [40,41] and ($\lambda_\mathrm{ij}-\lambda_\mathrm{ii}$) the interaction parameters to be determined. The NRTL equation for $G_\mathrm{m}^\mathrm{E}$ is:

$$G_\mathrm{m}^\mathrm{E}/RT = x_1x_2\left[\frac{\tau_{21}G_{21}}{x_1 + x_2G_{21}} + \frac{\tau_{12}G_{12}}{x_2 + x_1G_{21}}\right] \qquad (3)$$

here, $G_{ij} = \exp(-\alpha_{ij}\tau_{ij})$ with $\tau_{ij} = \frac{g_{ij}-g_{jj}}{RT}$. Along calculations, the value $\alpha_{ij} = \alpha_{ji} = 0.47$ has been used in order to take into account the non-randomness of the mixture. The interaction parameters ($g_{ij}-g_{jj}$) are adjusted to the experimental data.

### *3.2 DISQUAC*

Some relevant features of the model are shortly summarized. (i) DISQUAC is based on the rigid lattice theory developed by Guggenheim [42]. (ii) The total molecular volumes, $r_\mathrm{i}$, surfaces, $q_\mathrm{i}$, and the molecular surface fractions, $\alpha_\mathrm{i}$, of the mixture components are calculated additively using the group volumes $R_\mathrm{G}$ and surfaces $Q_\mathrm{G}$ recommended by Bondi [43]. The volume $R_\mathrm{CH4}$ and surface $Q_\mathrm{CH4}$ of methane are taken arbitrarily as volume and surface units [44]. The geometrical parameters for the groups considered along the work can be found in the literature [13,26]. (iii) The molar excess functions $G_\mathrm{m}^\mathrm{E}$ and $H_\mathrm{m}^\mathrm{E}$ (enthalpy) are determined by adding two contributions: a dispersive (DIS) term, related to the dispersive interactions; and a quasichemical (QUAC) term which arises from the anisotropy of the field forces created by the solution molecules. For $G_\mathrm{m}^\mathrm{E}$ a combinatorial term, $G_\mathrm{m}^\mathrm{E,COMB}$, represented by the Flory-Huggins equation [44,45] must be included. Thus,

$$G_{\mathrm{m}}^{\mathrm{E}} = G_{\mathrm{m}}^{\mathrm{E,COMB}} + G_{\mathrm{m}}^{\mathrm{E,DIS}} + G_{\mathrm{m}}^{\mathrm{E,QUAC}} \quad (4)$$

$$H_{\mathrm{m}}^{\mathrm{E}} = H_{\mathrm{m}}^{\mathrm{E,DIS}} + H_{\mathrm{m}}^{\mathrm{E,QUAC}} \quad (5)$$

For a binary mixture ($\mathrm{i} \neq \mathrm{j} = 1, 2$), the combinatorial contribution to the activity coefficients are determined from the expression:

$$\ln \gamma_{\mathrm{i}}^{\mathrm{COMB}} = \ln \frac{\varphi_{\mathrm{i}}}{x_{\mathrm{i}}} - x_{\mathrm{j}} \frac{\varphi_{\mathrm{i}}}{x_{\mathrm{i}}} + \varphi_{\mathrm{j}} \quad (6)$$

being $\varphi_{\mathrm{i}}$ the volume fraction, $\varphi_{\mathrm{i}} = \frac{x_{\mathrm{i}} r_{\mathrm{i}}}{\sum x_{\mathrm{j}} r_{\mathrm{j}}}$. (iv) It is assumed that the interaction parameters are dependent on the molecular structure; (v) For all the polar contacts, the same coordination number ($z$ = 4) is used. This is one of the more important shortcomings of the model, partially removed considering structure dependent interaction parameters. (vi) It is also assumed that there is no volume change upon mixing, i.e., the molar excess molar volume, $V_{\mathrm{m}}^{\mathrm{E}}$, is 0.

The equations used to calculate the DIS and QUAC contributions to $G_{\mathrm{m}}^{\mathrm{E}}$ and $H_{\mathrm{m}}^{\mathrm{E}}$ can be found elsewhere [24,46]. The temperature dependence of the interaction parameters is expressed in terms of the DIS and QUAC interchange coefficients [24,46], $C_{\mathrm{st,l}}^{\mathrm{DIS}}; C_{\mathrm{st,l}}^{\mathrm{QUAC}}$ where s≠t and l = 1 (Gibbs energy; $C_{\mathrm{st,1}}^{\mathrm{DIS/QUAC}} = g_{\mathrm{st}}^{\mathrm{DIS/QUAC}}(T_0)/RT_0$); l = 2 (excess enthalpy; $C_{\mathrm{st,2}}^{\mathrm{DIS/QUAC}} = h_{\mathrm{st}}^{\mathrm{DIS/QUAC}}(T_0)/RT_0$)), l = 3 (heat capacity; $C_{\mathrm{st,3}}^{\mathrm{DIS/QUAC}} = c_{\mathrm{pst}}^{\mathrm{DIS/QUAC}}(T_0)/R$)). $T_{\mathrm{o}}$ = 298.15 K is the scaling temperature. The corresponding equations are:

$$\frac{g_{\mathrm{st}}^{\mathrm{DIS/QUAC}}}{RT} = C_{\mathrm{st,1}}^{\mathrm{DIS/QUAC}} + C_{\mathrm{st,2}}^{\mathrm{DIS/QUAC}} \left[ \frac{T_0}{T} - 1 \right] + C_{\mathrm{st,3}}^{\mathrm{DIS/QUAC}} \left[ \ln\left(\frac{T_0}{T}\right) - \frac{T_0}{T} + 1 \right] \quad (7)$$

$$\frac{h_{\mathrm{st}}^{\mathrm{DIS/QUAC}}}{RT} = C_{\mathrm{st,2}}^{\mathrm{DIS/QUAC}} \frac{T_{\mathrm{o}}}{T} - C_{\mathrm{st,3}}^{\mathrm{DIS/QUAC}} \left( \frac{T_{\mathrm{o}}}{T} - 1 \right) \quad (8)$$

$$\frac{c_{\mathrm{pst}}^{\mathrm{DIS/QUAC}}}{R} = C_{\mathrm{st,3}}^{\mathrm{DIS/QUAC}} \quad (9)$$

### *3.3 Modified UNIFAC (Dortmund version)*

This version of UNIFAC [22] differs from the original UNIFAC [47] by the combinatorial term and the temperature dependence of the group interaction parameters $\psi_{\mathrm{nm}}$. This dependence is as follows:

$$\psi_{\rm nm} = \exp-[\frac{a_{\rm nm} + b_{\rm nm}T + c_{\rm nm}T^2}{T}] \quad (10)$$

where ( $a_{\rm nm}, b_{\rm nm}, c_{\rm nm}$ ) are the interaction parameters. A similar expression is valid for $\psi_{\rm mn}$ The equations used to calculate $G_{\rm m}^{\rm E}$ and $H_{\rm m}^{\rm E}$ are obtained from the well-known fundamental equation for $\gamma_i$ :

$$\ln\gamma_{\rm i} = \ln\gamma_{\rm i}^{\rm COMB} + \ln\gamma_{\rm i}^{\rm RES} \quad (11)$$

where $\ln\gamma_{\rm i}^{\rm COMB}$ and $\ln\gamma_{\rm i}^{\rm RES}$ represent the combinatorial and residual term, respectively.
The expression for the combinatorial part is:

$$\ln\gamma_{\rm i}^{\rm COMB} = 1 - \varphi_{\rm i}^{'} + \ln\varphi_{\rm i}^{'} - 5q_{\rm i}(1 - \frac{\varphi_{\rm i}}{\xi_{\rm i}} + \ln\frac{\varphi_{\rm i}}{\xi_{\rm i}}) \quad (12)$$

in this equation, $\varphi_{\rm i}^{'} = \frac{r_{\rm i}^{3/4}}{\sum x_{\rm j} r_{\rm j}^{3/4}}$ and $\xi_{\rm i} = \frac{x_{\rm i}q_{\rm i}}{x_1q_1 + x_2q_2}$ is the surface fraction.

For the residual part, the needed equations can be found elsewhere [46]. In UNIFAC (Dortmund), two main groups, OH and $CH_3OH$, are defined for predicting thermodynamic properties of mixtures with alkanols. The main group OH (Nº 5) is subdivided in three subgroups: OH(p), OH(s) and OH(t) for the representation of primary, secondary and tertiary alkanols, respectively. The $CH_3OH$ group (Nº 6) is a specific group for methanol solutions. The aromatic molecules considered in this work are characterized by the main group ACH (Nº 3), which is subdivided in two subgroups, ACH and AC. The subgroups within the same main group have different geometrical parameters, and identical group energy-interaction parameters. It must be mentioned that the geometrical parameters, the relative van der Waals volumes and the relative van der Waals surfaces are not calculated from molecular parameters like in the original UNIFAC, but fitted together with the interaction parameters to the experimental values of the thermodynamic properties considered. The geometrical and interaction parameters were taken from the literature and used without modifications [25].

*3.4 The concentration-concentration structure factor formalism*

For binary systems, the $S_{\rm CC}(0)$ function can be determined from the expression [28,29]:

$$S_{\mathrm{CC}}(0) = \frac{x_1 x_2}{1 + \frac{x_1 x_2}{RT}\left(\frac{\partial^2 G_{\mathrm{m}}^{\mathrm{E}}}{\partial x_1^2}\right)_{P,T}} \quad (13)$$

In the case of an ideal system, $G_{\mathrm{m}}^{\mathrm{E,id}}$ = 0; and $S_{\mathrm{CC}}^{\mathrm{id}}(0)$ = $x_1x_2$. Stability conditions require that $S_{\mathrm{CC}}(0)$ > 0. Thus, if a mixture is close to phase separation, $S_{\mathrm{CC}}(0)$ must be large and positive, the dominant trend is then the separation between components (homocoordination), and $S_{\mathrm{CC}}(0) > x_1x_2$. If compound formation between components is dominant (heterocoordination), $S_{\mathrm{CC}}(0)$ must be very low and $0 < S_{\mathrm{CC}}(0) < x_1x_2$ . Additional details are available in reference [28].

## 4. Determination of adjustable parameters

### *4.1 Wilson and NRTL parameters*

The ($\lambda_{\mathrm{ij}} - \lambda_{\mathrm{ii}}$) and ($g_{ij} - g_{jj}$) parameters were obtained through a Marquardt algorithm [48] which minimizes the function:

$$\Delta(T_{\mathrm{SLE}})/K = \frac{1}{N}\sum\left|T_{\mathrm{SLE,exp}} - T_{\mathrm{SLE,calc}}\right| \quad (14)$$

where $N$ stands for the number of experimental points. Values of the parameters are listed in Table 5.

**TABLE 5 goes here**

### *4.2 DISQUAC interaction parameters*

In terms of DISQUAC, the mixtures under study are regarded as possessing three surfaces: (i) type a, aliphatic, ($CH_3$, $CH_2$, in *n*-alkanes, or 1-alkanols); (ii) type b, aromatic in biphenyl or naphthalene; type h, OH in 1-alkanols). Therefore, the systems are characterized by the contacts: (a,b), (a,h) and (b,h). The interaction parameters for the (a,b) contacts are purely dispersive and have been determined from thermodynamic data for naphthalene or biphenyl + *n*-alkane systems [26,49]. The $C_{\mathrm{ah,l}}^{\mathrm{DIS}}$ and $C_{\mathrm{ah,l}}^{\mathrm{QUAC}}$ (l =1,2,3) coefficients are also known from the corresponding investigation of 1-alkanol + *n*-alkane systems [50,51]. Our previous investigations on 1-alkanol + naphthalene, or + biphenyl systems [13,26] show that the QUAC interaction parameters for the (b,h) contacts are the same as those given for 1-alkanol + benzene, or + toluene systems [27]. Here, we have applied the same rule and, under the assumption that the $C_{\mathrm{bh,l}}^{\mathrm{DIS}}$ (l =2,3) parameters are the same as in systems with benzene, we have determined the corresponding $C_{\mathrm{bh,l}}^{\mathrm{DIS}}$ value (Table 6), following the general procedure explained in detail elsewhere [24,46].

**TABLE 6 goes here**

## 5. Theoretical results

A comparison between SLE calculations using DISQUAC and UNIFAC models with experimental values is shown in Tables 7 and 8 (Figures. 1,2,3,4), which also include results from the Ideal Solubility Model. For the sake of clarity, relative deviations for the equilibrium temperature ($T_{SLE}$), defined as:

$$\sigma_r(T_{SLE}) = \left\{ \frac{1}{N} \sum \left[ \frac{T_{SLE,exp} - T_{SLE,calc}}{T_{SLE,exp}} \right]^2 \right\}^{1/2} \tag{15}$$

are given in Table 7, together with the corresponding mean absolute deviation (equation 14). The coordinates of the eutectic points, obtained from the different models applied, are listed in Table 8. UNIFAC calculations remain unchanged when a different set of interaction parameters [52] is used. A short comparison between experimental results for $S_{CC}(0)$ and DISQUAC values is presented in Figures 7 and 8.

**TABLES 7 and 8 goes here**

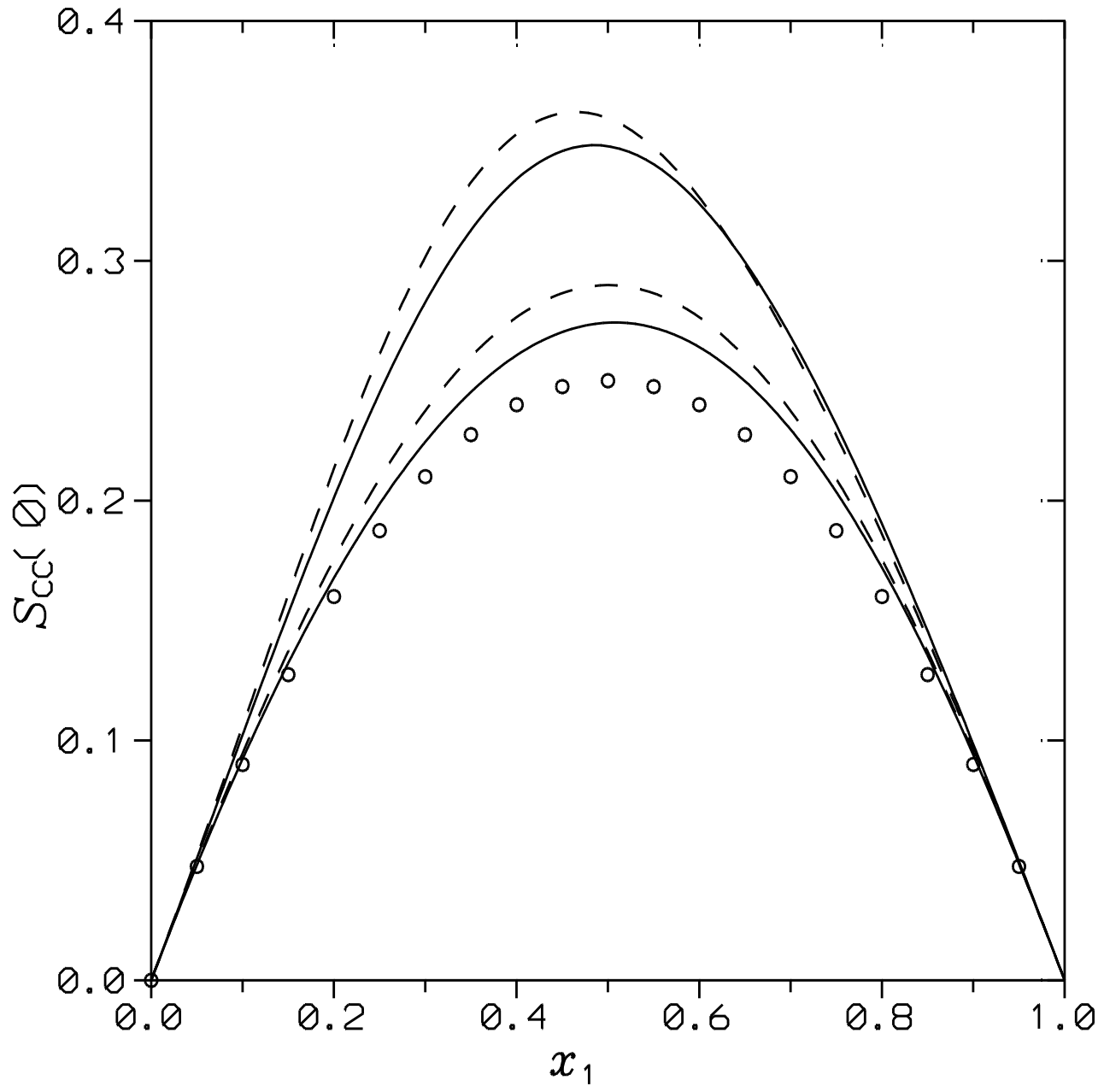

**Figure 7** $S_{CC}(0)$ of the naphthalene(1) (upper curves) or benzene(1) (lower curves) + cyclohexane(2) mixtures at 413.15 K. Dashed lines, experimental values [68,69]. Solid lines, DISQUAC calculations. Points, results for the ideal system

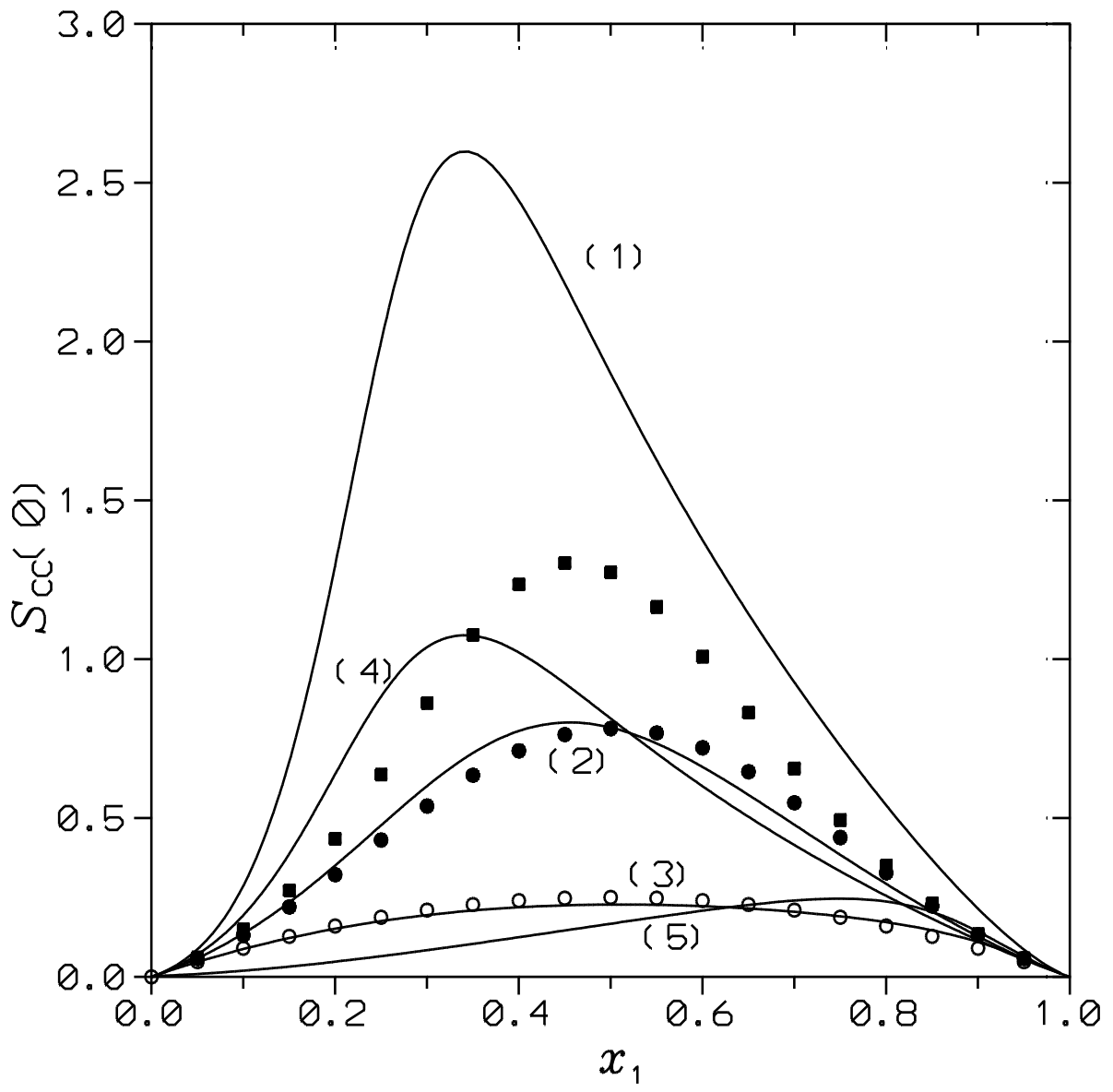

**Figure 8** $S_{CC}(0)$ of organic compound(1) + 1-alkanol(2) mixtures at temperature *T*. Lines, DISQUAC calculations. (1): heptane(1) + methanol(2) at 373.15 K, (2): benzene(1) + methanol(2) at 413.15 K; (3) benzene(1) + 1-hexadecanol(2) at 413.15 K. (4): naphthalene(1) + methanol(2) at 413.15 K. (5): naphthalene(1) + 1-hexadecanol(2) at 413.15 K. Full points, experimental results for benzene(1) + methanol(2) systems [70]: (■), *T* = 373.15 K; (●), *T* = 413.15 K. Open symbols, results for the ideal mixture

The model is a reliable tool for this type of calculations. For naphthalene + 1-alkanol mixtures, Figure 9 shows the maxima values of $S_{CC}(0)$ determined using DISQUAC.

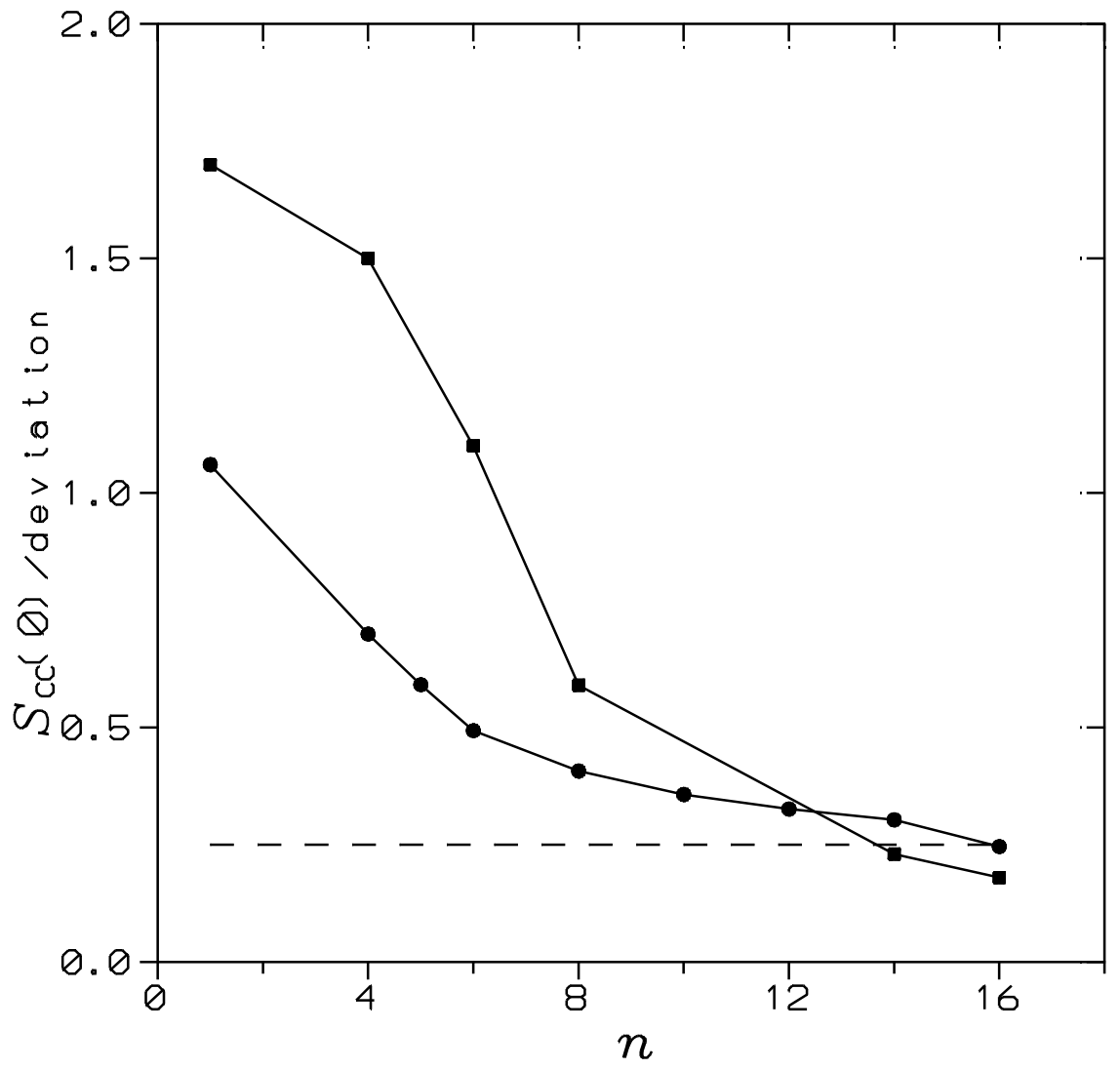

**Figure 9** $S_{CC}(0)$ and $100\sigma_r(T)$ (deviation) values for naphthalene(1) + 1-alkanol(2) mixtures vs. *n*, the number of carbon atoms of the 1-alkanol: (●), maxima of the $S_{CC}(0)$ curves at 413.15 K (this work); (■), ($100\sigma_r(T)$) (Table 7]. Lines are for the aid of the eye. Dashed line, $S_{CC}(0)$ results for the ideal model

## 6. Discussion

Below, $n$ stands for the number of C atoms in the 1-alkanol.

Firstly, we note that this type of systems is mainly characterized by positive deviations from Raoult's law (Table 3, Figure S3). For mixtures containing naphthalene, $\sigma_r(T_{\text{SLE}})$ values provided by the Ideal Solubility Model decrease when $n$ increases (Table 7). Thus, we have for the mixture with $n$ = 1, $\sigma_r(T_{\text{SLE}})$ = 0.17, $\Delta(T_{\text{SLE}})$ = 45 K, and for the solution with $n$ = 16, $\sigma_r(T_{\text{SLE}})$ = 0.018, $\Delta(T_{\text{SLE}})$ = 4.3 K. Similarly, for biphenyl systems, $\sigma_r(T_{\text{SLE}})$ = 0.030, $\Delta(T_{\text{SLE}})$ = 7.3 K ($n$ = 14, this work) and $\sigma_r(T_{\text{SLE}})$ = 0.012, $\Delta(T_{\text{SLE}})$ = 3.1 K [13] ($n$ = 20). This may be due to: (i) lower self-association of the longer 1-alkanols and/or (ii) size effects. These ones can be examined calculating the $G_{\text{m}}^{\text{E,COMB}}$ contribution using, e.g., the Flory-Huggins term. At 373.15 K, results are $-568$ J mol$^{-1}$ ($n$ = 1) and $-266$ J mol$^{-1}$ ($n$ = 16). This indicates that size effects are more relevant in the methanol mixture, and that the larger positive deviations from the Raoult's law of this system can be then ascribed to the larger self-association of methanol. The maxima values of $S_{\text{CC}}(0)$ (Figure 9) also decrease when $n$ increases, and this confirms that deviations from the ideal behaviour decrease when homocoordination becomes weaker.

Next, we analyse $S_{\text{CC}}(0)$ results. In the case of mixtures with cyclohexane ($T$ = 413.15 K), $S_{\text{CC}}(0)$ (naphthalene) > $S_{\text{CC}}(0)$ (benzene) (Figure 7), i.e., homocoordination is higher in the system with naphthalene. For the heptane + methanol mixture at 373.15 K, the $S_{\text{CC}}(0)$ curve shows a large maximum and it is skewed at higher mole fractions of the alcohol (Figure 8). The former is a consequence of the proximity of the upper critical solution temperature (324.5 K [53]). The asymmetry of the curve merely shows the large self-association of methanol. At the same temperature, the $S_{\text{CC}}(0)$ curve of the benzene + methanol mixture shows a lower maximum, since aromatic compounds are better breakers than alkanes of the alcohol network. Note that benzene and methanol are completely miscible at 298.15 K. For this system, $S_{\text{CC}}(0)$ decreases when the temperature is increased as the corresponding results at 413.15 K reveal (Figure 8). This is the normal behaviour since homocoordination decreases when $T$ increases. The replacement of methanol by 1-hexadecanol leads to lower $S_{\text{CC}}(0)$ results, which is consistent with the lower self-association of this alkanol. At 413.15 K, with regards to mixtures including naphthalene two statements can be given: i) for methanol mixtures, $S_{\text{CC}}(0)$ (naphthalene) > $S_{\text{CC}}(0)$ (benzene) and the naphthalene curve is more shifted to higher molar fractions of methanol. This seems to indicate that naphthalene is a poorer breaker of the methanol self-association. (ii) For the naphthalene + 1-hexadecanol system, the $S_{CC}(0)$ curve is skewed to higher mole fractions of the aromatic compound, which reveals that, in the mentioned region, interactions between naphthalene molecules are more relevant. At 413.15 K, the maxima values of $S_{CC}(0)$ for mixtures with biphenyl are: 0.42 ($n$ = 14) and 0.26 ($n$ = 16), which are close to those of the corresponding solutions with naphthalene: 0.30 and 0.25, respectively. Thus, for this type of systems, the model predicts no meaningful difference regarding their homocoordination.

### *6.1 Comparison between DISQUAC and UNIFAC results*

For the systems measured in this work, both models provide rather similar results (Tables 7 and 8), although DISQUAC gives a slightly better overall description of the SLE curves (Figures 1-4). In order to attain a better understanding of the theoretical results, we have also considered the systems naphthalene + methanol, or + 1-butanol, or + 1-hexanol, or + 1-octanol, and proceeded as follows. Firstly, we have examined the impact of the combinatorial term on the mentioned results. At this end, we have conducted calculations, for the selected mixtures, neglecting the interactional contribution to the activity coefficients. i.e., we have assumed that activity coefficients are merely described by equations (6) or (12). In such a case, we note that that the differences between experimental and theoretical results are lower when equation (12) is employed, that is, size effects are better represented by equation (12), while the Flory-Huggins term overestimates such effects (Table 7). The observed differences between DISQUAC and UNIFAC results may be then due to: i) a better temperature dependence of the interaction parameters in the case of DISQUAC; (ii) it is assumed that the DISQUAC interaction parameters depend on the molecular structure. The latter may explain the better results obtained for the solutions with 1-butanol, 1-hexanol, or 1-octanol. However, SLE data of the naphthalene + methanol mixture are also better described by DISQUAC (Table 7). This is remarkable due to: (a) the range of measured temperatures is rather wide: (274-353.37) K; (b) methanol is a main group within UNIFAC. This seems to indicate that the temperature dependence of the DISQUAC interactions parameters is more suitable. We remark that results from the Ideal Solubility Model for systems including the shorter 1-alkanols are much poorer, which clearly shows the large non-ideality of such solutions (Table 7).

Finally, we have compared DISQUAC and UNIFAC results on excess molar enthalpies, $H_m^E$, with the corresponding experimental value for the mixture naphthalene + $n$-$C_{28}$ at 391.75 K and equimolar composition: 1220 (DISQUAC); 1414 (UNIFAC), 1206 (experimental result [54], all values in J mol$^{-1}$). This suggests that, probably, a new main group should be defined for a better estimation of the thermodynamic properties of solutions including PAHs.

## 7. Conclusions

SLE data have been reported for the eutectic mixtures naphthalene or biphenyl + 1-tetradecanol, or + 1-hexadecanol. The systems show decreasing positive deviations from the Raoults's law when the alcohol size increases, i.e., when homocoordination becomes weaker. DISQUAC and UNIFAC describe similarly the phase diagrams of the measured mixtures. The comparison of theoretical results from both models for naphthalene + 1-alkanol mixtures shows that the temperature dependence of the interaction parameters is more suitable in DISQUAC.

### Author contributions

L.F. Sanz and F. Hevia performed the experimental work. Data correlation was conducted by J.A. González and L.F. Sanz. J.A. González, I. García de la Fuente and J.C. Cobos wrote the draft of the manuscript. F. Hevia prepared the Figures. L.F. Sanz and J.A. González wrote the final version of the manuscript, reviewed by all authors.

### Funding

This work was supported by Consejería de Educación de Castilla y León, under Project VA100G19 (Apoyo a GIR, BDNS: 425389.

### Declarations

Authors declare no competing financial interest, or of any type

## 8. References

[1] Farahani, B. V., Rajabi, F. H., Hosseindoust, B., Zenooz, N. DSC study of solid-liquid equilibria for energetic binary mixtures of methylnitramine with 2,4-dinitro-2,4-diazapentane and 2,4-dinitro-2,4-diazahexane. J. Phase Equilib. Diffus. **31**, 536–541 (2010).

[2] Ulrich, J., Bülau, H.C. Melt crystallization, in Handbook of Industrial Crystallization ($2^{nd}$ edition). A. S. Myerson, Editor. Butterworth-Heinemann, 161-179 (2002).

[3] Ahlers, J., Lohmann, J. Gmehling, J. Binary solid-liquid equilibria of organic systems containing different amides and sulfolane. J. Chem. Eng. Data **44**, 727-730 (1999).

[4] Wittig, R., Constantinescu, D., Gmehling, J. Binary solid-liquid equilibria of organic systems containing $\varepsilon$-caprolactone. J. Chem. Eng. Data **46**, 1490-1493 (2001).

[5] Sarbu, I., Sebarchievici, C. A. Comprehensive review of thermal energy storage. Sustain. **10**, 191 (2018).

[6] Fallahi, A., Guldentops, G., Tao, M., Granados-Focil, S., Van Dessel, S. Review on solid-solid phase change materials for thermal energy storage: molecular structure and thermal properties. Appl. Therm. Eng. **127**, 1427-1441 (2017).

[7] Abbott, A. P., Boothby, D., Capper, G., Davies, D. L., Rasheed, R. K.. Deep eutectic solvents formed between choline chloride and carboxylic acids: versatile alternatives to ionic liquids. J. Am. Chem. Soc. **126**, 9142-9147 (2004)

[8] Abbott, A. P., Capper, G., Davies, D. L., Rasheed, R. K., Tambyrajah, V. Novel solvent properties of choline chloride/urea mixtures. Chem. Commun. 70-71 (2003)

[9] Foreman, M.R.S. Progress towards a process for the recycling of nickel metal hydride electric cells using a deep eutectic solvent. Cogent Chem. **2**, 1139289–1139300 (2016).

[10] Łukomska, A., Wiśniewska, A., Dąbrowski, Z., Kolasa, D., Luchcińska, S., Lach, J., Wróbel, K., Domańska, U, Recovery of zinc and manganese from ”black mass” of waste Zn-MnO2 alkaline batteries by solvent extraction technique with ionic liquids, DES and organophosphorous-based acids, J. Mol. Liq. **338**, 116590 (2021).

[11] Domanska, U. Solubility and hydrogen bonding. Part VII. Synergic effect of solubility of naphthalene in mixed solvents. Polish. J. Chem. **55**, 1715-1720 (1981).

[12] Ward, H.L. The solubility relations of naphthalene. J. Phys. Chem. **30**, 1316-1333 (1926).

[13] Boudouh, I., González, J.A., Djemani, I., Barkat, D. Solid-liquid equilibria of eicosane, tetracosane, or biphenyl + 1-octadecanol, or + 1-eicosanol mixtures. Fluid Phase Equilib. **442**, 28-37 (2017).

[14] Hawes, D.W., Feldman, D., Banu, D. Latent heat storage in building materials, Energy Build. **20**, 77-86 (1993).

[15] Burley, S.K., Petsko, G.A. Weakly polar interactions in proteins. Adv. Protein Chem. **39**, 125-189 (1988).

[16] Cabaleiro, D., Gracia-Fernández, C., Lugo, L. (Solid-liquid) phase equilibria and heat capacity of (diphenyl ether + biphenyl) mixtures used as thermal energy storage materials. J. Chem. Thermodyn. **74**, 43-50 (2014).

[17] Cabaleiro, D., Pastoriza-Gallego, M.J., Piñeiro, M.M., Legido, J.L., Lugo, L. Thermophysical properties of (diphenyl ether + biphenyl) mixtures for their use as heat transfer fluids. J. Chem. Thermodyn. **50**, 80–88 (2012)

[18] Demirbas, A. Asphaltene yields from five types of fuels via different methods. Energy Conversion Manag. **43**, 1091-1097 (2002).

[19] Silva, S.M.C., Rajagopal, K. Steady state size distribution of asphaltenes by flocculation from toluene-*n*-heptane mixtures. Petrol. Sci. Technol. **22**, 1073-1085 (2004).

[20] Wilson, G.M. Vapor-liquid equilibrium. XI. A new expression for the excess free energy of mixing. J. Am. Chem. Soc. **86**, 127-130 (1964).

[21] Renon, H., Prausnitz, J. M. Local compositions in thermodynamic excess functions for liquid mixtures. AIChE J. **14**, 135-144 (1968).

[22] Weidlich, U., Gmehling, J. A modified UNIFAC model. 1. Prediction of VLE, $H^{\mathrm{E}}$ and $\gamma^{\infty}$ Ind. Eng. Chem. Res. **26**, 1372-1381 (1987).

[23] Kehiaian, H.V. Group contribution methods for liquid mixtures: a critical review. Fluid Phase Equilib. **13**, 243-252 (1983).

[24] González, J.A., García de la Fuente, I., Cobos, J.C. Correlation and prediction of excess molar enthalpies using DISQUAC in: E. Wilhelm, T.M. Letcher (Eds.), Enthalpy and Internal Energy: Liquids, Solutions and Vapours, Royal Society of Chemistry, Croydon (2017).

[25] Gmehling, J., Li, J., Schiller, M. A modified UNIFAC model, 2. Present parameter matrix and results for different thermodynamic properties. Ind. Eng. Chem. Res. 32 (1993) 178-193.

[26] Gonzalez, J.A., Zawadzki, M., Domanska, U. Thermodynamics of mixtures containing polycyclic aromatic hydrocarbons. J. Mol. Liq. **143**, 134-140 (2008).

[27] Domanska, U. González, J.A. Solid-liquid equilibria for systems containing long-chain 1-alkanols. I. Experimental data for 1-dodecanol, 1-tetradecanol, 1-hexadecanol, 1-octadecanol or 1-icosanol + benzene, or toluene mixtures. Characterization in terms of DISQUAC. Fluid Phase Equilib. **119**, 131-151 (1996).

[28] Cobos, J.C. An exact quasi-chemical equation for excess heat capacity with W-shaped concentration dependence. Fluid Phase Equilib. **133**, 105-127 (1997).

[29] Trejo, L.M., Costas, M., Patterson, D.. Effect of molecular size on the W-shaped excess heat capacities: oxaalkane-alkane systems. J. Chem. Soc., Faraday Trans. **87**, 3001-3008 (1991).

[30] CIAAW, Atomic weights of the elements 2015, ciaaw.org/atomic-weights.htm, (2015).

[31] Stolen S., Gronvold, F. Critical assessment of the enthalpy of fusion of metals used as enthalpy standards at moderate to high temperatures. Thermochim. Acta **327**, 1-32 (1999).

[32] Mosselman, C., Mourik., Dekker, H. Enthalpies of phase change and heat capacities of some long-chain alcohols. Adiabatic semi-microcalorimeter for studies of polymorphism. J. Chem. Thermodyn. **6**, 477-487 (1974).

[33] Metivaud, V., Lefevre, A., Ventola, L., Negrier, P., Moreno, E., Calvet, T., Mondieig, D., Cuevas-Diarte, M.A., Hexadecane ($C_{16}H_{34}$) + 1-hexadecanol ($C_{16}H_{33}OH$) binary

system: crystal structures of the components and experimental phase diagram. Application to thermal protection of liquids, Chem. Mater. **17**, 3302–3310 (2005).

[34] Kuchhal, Y.K., Shukla, R.N., Biswas, A.B. Differential thermal-analysis of *n*-long chain alcohols and corresponding alkoxy ethanols, Thermochim. Acta **31**, 61-70 (1979).

[35] Berchiesi, G. Int. DATA Ser. Selec. Data Mixtures, Ser. A **2**, 95-100 (1985).

[36] Prausnitz, J.M., Lichtenthaler, R.N., Gomes de Azevedo, E. Termodinámica molecular de los equilibrios entre fases. Prentice-Hall, Madrid (2000).

[37] Boudouh, I., Hafsaoui,S.L., Mahmoud, R., Barkat, D. Measurement and prediction of solid-phase equilibria for systems containing biphenyl in binary solution with long-chain *n*-alkanes. J. Therm. Anal. Calorim. **125**, 793-801 (2016).

[38] Costa, M.C., Rolemberg, M.-P., Meirelles, A.J.A., Coutinho, J.P.A., Krähenbühl, M.A. The solid-liquid phase diagrams of binary mixtures of even saturated fatty acids differeing by six carbon atoms. Thermochim. Acta **496**, 30-37 (2009).

[39] Inoue, T., Hisatsugu, Y., Ishikawa, R., Suzuki, M. Solid-liquid phase behaviour of binary fatty acid mixtures: 2. Mixtures of oleic acid with lauric acid, myristic acid, and palmitic acid. Chem. Phys. Lipids **127**, 161-173 (2004).

[40] Grzyll, L. R., Ramos, C., Back, D. D. Density, viscosity, and surface tension of liquid quinoline, naphthalene, biphenyl, decafluorobiphenyl, and 1,2-diphenylbenzene from 300 to 400 ºC. J. Chem. Eng. Data, **41**, 446-450 (1996).

[41] Liew, K. Y., Seng, C. E., Ng, B. H. Molar volumes of *n*-alcohols from 15 to 80 ºC J. Solution Chem., 21 1177-1183 (1992).

[42] Guggenheim, E.A. Mixtures, Oxford University Press, Oxford (1952).

[43] Bondi, A. Physical Properties of Molecular Crystals, Liquids and Glasses, Wiley, New York (1968).

[44] Kehiaian, H.V., Grolier, J.-P.E., Benson, G.C. Thermodynamics of organic mixtures. A generalized quasichemical model theory in terms of group surface interactions. J. Chim. Phys. **75**, 1031-1048 (1978).

[45] González, J.A., García de la Fuente, I., Cobos, J.C., Casanova, C., Ait-Kaci, A. Application of the zeroth approximation of the DISQUAC model to cyclohexane + *n*-alkane mixtures using different combinatorial entropy terms. Fluid Phase Equilib. **112**, 63-87 (1995).

[46] González, J.A., García de la Fuente, I., Cobos, J.C. Thermodynamics of mixtures with strongly negative deviations from Raoult's law. Part 4. Application of the DISQUAC model to mixtures of 1-alkanols with primary or secondary linear amines. Comparison with Dortmund UNIFAC and ERAS results. Fluid Phase Equilib. **168**, 31-58 (2000).

[47] Fredenslund, A., Jones, R.L., Prausnitz, J.M. Group-contribution estimation of activity coefficients in nonidel liquid mixtures. AIChE J. **21**, 1086-1099 (1975).

[48] Bevington, P.R. Data Reduction and Error Analysis for the Physical Sciences, McGraw-Hill, New York, (1969).

[49] Boudouh, I., Djemai, I., González, J.A., Barkat, D. Solid-liquid equilibria of biphenyl binary systems J. Mol. Liq. **216**, 764-770 (2016).

[50] González, J.A., García de la Fuente, I., Cobos, J.C., Casanova, C. A characterization of the aliphatic/hydroxyl interactions using a group contribution model (DISQUAC). Ber. Bunsenges. Phys. Chem. **95**, 1658-1668 (1991).

[51] González, J.A., García de la Fuente, I., Cobos, J.C., Casanova, C., Domanska, U. DISQUAC application to SLE of binary mixtures containing long chain 1-alkanols (1-tetradecanol, 1-hexadecanol, 1-octadecanol, or 1-eicosanol) and *n*-alkanes ($C_8$-$C_{16}$). Ber. Bunsenges. Phys. Chem. **98**, 955-959 (1994).

[52] Kang, J.W., Diky, V., Frenkel, M. New modified UNIFAC parameters using critically evaluated phase equilibrium data. Fluid Phase Equilib. **388**, 128–141 (2015).

[53] Ott, J. B., Holscher, I. F., Schneider, G. M. (Liquid + liquid) phase equilibria in (methanol + heptane) and (methanol + octane) at pressures from 0.1 to 150 MPa J. Chem. Thermodyn. **18**, 815-826 (1986).

[54] Aoulmi, A., Bouroukba, M., Solimando, R., Rogalski, M. Thermodynamics of mixtures formed by polycyclic aromatic hydrocarbons with long chain alkanes. Fluid Phase Equilib. **110**, 283-297 (1995).

[55] Coon, J.E., Sediawan, W.B., Auwaerter, J.E., McLaughlin, E. Solubilities of families of heterocyclic polynuclear aromatics in organic solvents and their mixtures, J. Solution Chem. **17**, 519–534 (1988).

[56] Chirico, R.D., Knipmeyer, S.E., Nguyen, A., Steele, W. V. The thermodynamic properties of biphenyl. J. Chem. Thermodyn. **21**, 1307-1331 (1989).

[57] Leys, J., Losada-Pérez, P., Slenders, E., Glorieux, C., Thoen, J. Investigation of the melting behaviour of the reference materials biphenyl and phenyl salicytate by a new type adiabatic scanning calorimeter. Thermochim. Acta **582**, 68-76 (2014).

[58] Sharma, K.P., Rai, R.N. Synthesis and characterization of novel binary organic monotectic and eutectic alloys. Thermochim. Acta **535**, 66-70 (2012).

[59] Chirico, R.D., Knipmeyer, S.E., Steele, W. V. Heat capacities, enthalpy increments, and derived thermodynamic functions for naphthalene between the temperatures 5 K and 440 K. J. Chem. Thermodyn. **34**, 1874-1884 (2002)

[60] Khimeche, K., Dahmani, A. Solid-liquid equilibria of naphthalene + alkanediamine mixtures. J. Chem. Eng. Data **51**, 383-385 (2006).

[61] Acree, W.E. Thermodynamic properties of organic compounds: enthalpy of fusion and melting point temperature compilation. Thermochim Acta **189**, 37-56 (1991).

[62] Sharma, B.L., Gupta, S., Tandon, S., Kant, R. Physico-mechanical properties of naphthalene-acenaphthene eutectic system by different modes of solidification, Materials Chemistry and Physics, **111**, 423-430 (2008).

[63] Maximo, G.J., Carareto, N. D. D., Costa, M. C., Dos, Santos, A. O., Cardoso, L.P., Krähenbühl M.A., Meirelles, A.J.A. On the solid-iquid equilibrium of binary mixtures of fatty alcohols and fatty acids. Fluid Phase Equilib. **366**, 88-98 (2014).

[64] Tian, T., Song, J., Niu, L., Feng, R. Preparation and properties of 1-tetradecanol/1,3:2,4-di-(3,4-dimethyl) benzylidene sorbitol gelatinous form-stable phase change materials. Thermochim. Acta **554**, 54-58 (2013).

[65] Zuo, J., Li, W., Weng, L. Thermal properties of lauric acid/1-tetradecanol binary system for energy storage. App. Thermal Eng. **31**, 1352-1355 (2011).

[66] Davies, M., Kybett, B. Sublimation and vaporization heats of long-chain alcohols. Trans. Faraday Soc. 61, 1608-1617 (1965).

[67] Xing, J., Tan, Z.-C., Shi, Q., Tong, B., Wang, S.-X., Li, Y.-S. Heat capacity and thermodynamic properties of 1-hexadecanol. J. Themal Anal. Calorim. **92**, 375–380 (2008)

[68] Wiśniewska, B., Gregorowicz, J., Malanowski, S. Development of a vapour-liquid equilibrium apparatus to work at pressures up to 3 MPa. Fluid Phase Equilib. **86**, 173-186 (1993).

[69] Lee, C.H., Holder, G.D. Vapor-liquid equilibria in the systems toluene/naphthalene and cyclohexane/naphthalene. J. Chem. Eng. Data **38**, 320-323 (1993).

[70] Butcher, K.L., Medani, M.S. Thermodynamic properties of methanol-benzene mixtures at elevated temperatures J. Appl. Chem. **18**, 100-107 (1968).

**TABLE 1**

Sample description

| Chemical | CAS number | $M^a$ /g mol$^{-1}$ | Source | Purity$^b$ |
|---|---|---|---|---|
| biphenyl | 92-52-4 | 154.2078 | Sigma-Aldrich | > 0.999 |
| naphthalene | 91-20-3 | 128.1705 | Sigma-Aldrich | > 0.999 |
| 1-tetradecanol | 112-72-1 | 214.3930 | Merck | > 0.993 |
| 1-hexadecanol | 36653-82-4 | 242.440 | Sigma-Aldrich | > 0.999 |

$^a$Molar mass; $^b$in mole fraction, by gas chromatography provided by the supplier

**TABLE 2**

Physical properties[a] of pure compounds at 0.1 MPa: melting temperature, $T_{\mathrm{m}}$, enthalpy of fusion, $\Delta H_{\mathrm{m}}$; $\Delta C_{p\mathrm{m}}$, heat capacity change at the melting point.

| compound | $T_{\mathrm{m}}$ /K | | $\Delta H_{\mathrm{m}}$ / kJ mol$^{-1}$ | | $\Delta C_{p\mathrm{m}}$ /J mol$^{-1}$ K$^{-1}$ |
|---|---|---|---|---|---|
| | this work | literature | this work | literature | |
| biphenyl | 342.2 | 342.10 [16] | 18.9 | 18.54 [16] | 36.3 [55] |
| | | 342.098 [56] | | 18.57 [56] | |
| | | 342.08 [57] | | 18.6 [57] | |
| | | 343.5[58] | | 19.3[58] | |
| naphthalene | 353.4 | 353.37 [59] | 19.3 | 18.99 [59] | |
| | | 354.69 [60] | | 19.55 [60] | |
| | | 353.4[61] | | 19.1 [61,62] | |
| | | 353.5 [62] | | | |
| 1-tetradecanol | 310.8 | 311.39 [63] | 49.2 | 47.60 [63] | 122 [32] |
| | | 309.55 [64] | | 51.17 [64] | |
| | | 311.00 [32] | | 49.51 [32] | |
| | | 309.8 [65] | | 47.81 [65] | |
| | | 311 [66] | | 39.75 [66] | |
| 1-hexadecanol | 321.8 | 322.225 [67] | 55.8 | 57.743 [67] | 137.3 [67] |
| | | | | 58.38 [29] | |
| | | 322 [66] | | 57.74 [66] | |
| | | | | 56.4 [63] | |
| | | 322.2 [32] | | 58.38 [32] | |

[a]The standard uncertainty for pressure is $u(P)$ = 0.2 kPa; the combined expanded uncertainty (0.95 level of confidence) for temperatures is $U_{\mathrm{c}}(T_{\mathrm{m}}) =$ 0.6 K and the relative combined expanded uncertainty (0.95 level of confidence) for melting enthalpies is $U_{\mathrm{rc}}(\Delta H_{\mathrm{m}}) = 0.06$

**TABLE 3**

Solid-liquid equilibrium temperatures, $T_{SLE}$, and activity coefficients, $\gamma_i^{exp}$ , for naphthalene(1) or biphenyl(1) + 1-tetradecanol(2) or + 1-hexadecanol(2) mixtures at 0.1 MPa[a]

| $x_1$ | $T_{SLE}$/K | solid phase[b] | $\gamma_1^{exp}$ [d] | $\gamma_2^{exp}$ [d] |
|---|---|---|---|---|
| | | naphthalene(1) + 1-tetradecanol(2) | | |
| 0.0000 | 310.8 | Alkanol(cr,II) | | 1.000 |
| 0.0537 | 308.9 | Alkanol(cr,II) | | 0.946 |
| 0.1070 | 308.8 | Alkanol(cr,II) | | 0.995 |
| 0.208[c] | 306.8 | | | |
| 0.3120 | 317.2 | Naphthalene (cr,I) | 1.519 | |
| 0.3987 | 323.8 | Naphthalene (cr,I) | 1.378 | |
| 0.4924 | 329.9 | Naphthalene (cr,I) | 1.274 | |
| 0.6002 | 334.5 | Naphthalene (cr,I) | 1.150 | |
| 0.6979 | 339.9 | Naphthalene (cr,I) | 1.104 | |
| 0.7997 | 344.6 | Naphthalene (cr,I) | 1.058 | |
| 0.8999 | 348.0 | Naphthalene (cr,I) | 1.004 | |
| 1.0000 | 353.4 | Naphthalene (cr,I) | 1.000 | |
| | | naphthalene(1) + 1-hexadecanol(2) | | |
| 0.0000 | 321.8 | Alkanol(cr,II) | | 1.000 |
| 0.0511 | 320.0 | Alkanol(cr,II) | | 0.936 |
| 0.1040 | 319.9 | Alkanol(cr,II) | | 0.985 |
| 0.1579 | 319.5 | Alkanol(cr,II) | | 1.022 |
| 0.2043 | 318.8 | Alkanol(cr,II) | | 1.034 |
| 0.271[c] | 316.5 | | | |
| 0.3044 | 322.4 | Naphthalene (cr,I) | 1.752 | |
| 0.3990 | 326.6 | Naphthalene (cr,I) | 1.465 | |
| 0.5022 | 331.3 | Naphthalene (cr,I) | 1.288 | |
| 0.5999 | 334.8 | Naphthalene (cr,I) | 1.160 | |
| 0.6998 | 339.3 | Naphthalene (cr,I) | 1.088 | |
| 0.8013 | 344.4 | Naphthalene (cr,I) | 1.053 | |
| 0.9010 | 348.5 | Naphthalene (cr,I) | 1.013 | |
| 1.0000 | 353.4 | Naphthalene (cr,I) | 1.000 | |

TABLE 3 (continued)

| | | | | |
|---|---|---|---|---|
| | | biphenyl(1) + 1-tetradecanol(2) | | |
| 0.0000 | 310.8 | Alkanol(cr,II) | | 1.000 |
| 0.0513 | 309.3 | Alkanol(cr,II) | | 0.962 |
| 0.1009 | 308.9 | Alkanol(cr,II) | | 0.990 |
| 0.201[c] | 307.2 | | | |
| 0.3017 | 315.5 | Biphenyl (cr,I) | 1.885 | |
| 0.4008 | 320.9 | Biphenyl (cr,I) | 1.602 | |
| 0.5045 | 327.6 | Biphenyl (cr,I) | 1.472 | |
| 0.5969 | 329.8 | Biphenyl (cr,I) | 1.303 | |
| 0.6959 | 332.8 | Biphenyl (cr,I) | 1.189 | |
| 0.7987 | 335.4 | Biphenyl (cr,I) | 1.093 | |
| 0.8967 | 337.6 | Biphenyl (cr,I) | 1.018 | |
| 1.0000 | 342.2 | Biphenyl (cr,I) | 1.000 | |
| | | biphenyl(1) + 1-hexadecanol(2) | | |
| 0.0000 | 321.8 | Alkanol(cr,II) | | 1.000 |
| 0.0519 | 320.8 | Alkanol(cr,II) | | 0.987 |
| 0.0990 | 320.2 | Alkanol(cr,II) | | 0.999 |
| 0.1521 | 319.7 | Alkanol(cr,II) | | 1.026 |
| 0.1982 | 318.9 | Alkanol(cr,II) | | 1.032 |
| 0.267[c] | 316.8 | | | |
| 0.4031 | 322.4 | Biphenyl (cr,I) | 1.647 | |
| 0.5012 | 324.9 | Biphenyl (cr,I) | 1.400 | |
| 0.6052 | 329.6 | Biphenyl (cr,I) | 1.282 | |
| 0.6994 | 332.5 | Biphenyl (cr,I) | 1.176 | |
| 0.7873 | 334.8 | Biphenyl (cr,I) | 1.097 | |
| 0.8950 | 337.7 | Biphenyl (cr,I) | 1.022 | |
| 1.0000 | 342.2 | Biphenyl (cr,I) | 1.000 | |

[a]standard uncertainties are: $u(P)$ = 0.2 kPa; $u(x_1)$ = 0.0005, and the combined expanded uncertainty (0.95 level of confidence) for temperature is $U_c(T)$ = 0.6 K; [b](cr) describes a single solid phase; (cr,I) represents the solid phases I of naphthalene or biphenyl, and (cr,II) describes the solid phases of the considered 1-alkanols; [c] composition of the eutectic point determined from Tamman's plot (see Table 4 and Figures 5-6); [d]values determined using equation (1) with physical constants listed in Table 2.

**TABLE 4**

Heat of melting, $\Delta H_{\mathrm{m}}$ and eutectic heat, $\Delta H_{\mathrm{eu}}$, for naphthalene(1) or biphenyl(1) + 1-tetradecanol(2) or + 1-hexadecanol(2) mixtures at 0.1 MPa.[a]

| $x_1$ | $\Delta H_{\mathrm{eu}}$ / kJ mol$^{-1}$ | $\Delta H_{\mathrm{m}}$ / kJ mol$^{-1}$ |
|---|---|---|
| | naphthalene(1) + 1-tetradecanol(2) | |
| 0.0000 | | 49.2 |
| 0.0537 | 10.2 | 36.5 |
| 0.1070 | 24.4 | 19.9 |
| 0.2080 | 41.6 | 0.0 |
| 0.3120 | 36.3 | 0.6 |
| 0.3987 | 31.3 | 3.0 |
| 0.4924 | 27.0 | 5.0 |
| 0.6002 | 22.0 | 7.4 |
| 0.6979 | 16.2 | 10.8 |
| 0.7997 | 10.4 | 14.2 |
| 0.8999 | 3.1 | 16.7 |
| 1.0000 | | 19.3 |
| | naphthalene(1) + 1-hexadecanol(2) | |
| 0.0000 | | 55.8 |
| 0.0511 | 6.3 | 47.0 |
| 0.1040 | 15.3 | 36.2 |
| 0.1579 | 27.6 | 23.0 |
| 0.2043 | 37.4 | 11.0 |
| 0.2705 | 48.5 | 0 |
| 0.3044 | 44.7 | 0.1 |
| 0.3990 | 40.5 | 0.3 |
| 0.5022 | 34.0 | 4.8 |
| 0.5999 | 27.3 | 7.9 |
| 0.6998 | 20.4 | 10.8 |
| 0.8013 | 12.9 | 14.0 |
| 0.9010 | 6.4 | 16.3 |
| 1.0000 | | 19.3 |
| | Biphenyl(1) + 1-tetradecanol(2) | |
| 0.0000 | | 49.2 |

TABLE 4 (continued)

| | | |
|---|---|---|
| 0.0513 | 11.5 | 35.7 |
| 0.1009 | 20.8 | 25.8 |
| 0.2010 | 42.4 | 0 |
| 0.3017 | 38.6 | 1.7 |
| 0.4008 | 31.1 | 2.9 |
| 0.5045 | 24.4 | 6.8 |
| 0.5969 | 20.4 | 8.1 |
| 0.6959 | 15.7 | 11.4 |
| 0.7987 | 10.3 | 14.4 |
| 0.8967 | 5.5 | 16.4 |
| 1.0000 | | 18.9 |
| | biphenyl(1) + 1-hexadecanol(2) | |
| 0.0000 | | 55.8 |
| 0.0519 | 7.0 | 46.6 |
| 0.0990 | 15.6 | 37.6 |
| 0.1521 | 27.3 | 22.8 |
| 0.1982 | 39.0 | 10.4 |
| 0.2670 | 50.3 | 0 |
| 0.4031 | 42.1 | 0.6 |
| 0.5012 | 33.3 | 2.2 |
| 0.6052 | 26.0 | 5. |
| 0.6994 | 19.6 | 7.95 |
| 0.7873 | 13.9 | 11.3 |
| 0.8950 | 6.5 | 14.8 |
| 1.0000 | | 18.9 |

[a]standard uncertainties are: $u(P)$ = 0.2 kPa; $u(x_1)$ = 0.0005 and the combined expanded uncertainty (0.95 level of confidence) for enthalpy is $U_{\mathrm{rc}}(\Delta H_{\mathrm{m}}) = 0.06$

**TABLE 5**

Adjustable parameters of the Wilson[a], $(\lambda_{ij}-\lambda_{ii})$, and NRTL[b], $(g_{ij}-g_{jj})$, equations.

| $(\lambda_{12}-\lambda_{11})$/ J mol$^{-1}$ | $(\lambda_{21}-\lambda_{22})$/ J mol$^{-1}$ | $(g_{12}-g_{22})$/ J mol$^{-1}$ | $(g_{21}-g_{11})$ J mol$^{-1}$ | $\Delta(T)$ [c]/K | | $\sigma_r(T)$ [d] | |
|---|---|---|---|---|---|---|---|
| | | | | Wilson | NRTL | Wilson | NRTL |
| naphthalene(1) + 1-tetradecanol(2) | | | | | | | |
| 5800 | −1700 | −1800 | 600 | 0.71 | 0.75 | 0.003 | 0.003 |
| naphthalene(1) + 1-hexadecanol(2) | | | | | | | |
| 6400 | −1500 | −1200 | 2500 | 0.32 | 0.30 | 0.001 | 0.0001 |
| biphenyl(1) + 1-tetradecanol(2) | | | | | | | |
| 3900 | 700 | −1900 | 1700 | 1.20 | 1.73 | 0.005 | 0.008 |
| biphenyl(1) + 1-hexadecanol(2) | | | | | | | |
| 8700 | −400 | −1800 | 2700 | 0.94 | 1.09 | 0.003 | 0.005 |

[a] equation (2); [b]equation (3); [c]equation (14); [d]equation (15)

**TABLE 6**

Dispersive, $C^{\mathrm{DIS}}_{\mathrm{bh},l}$, and quasichemical, $C^{\mathrm{QUAC}}_{\mathrm{bh},l}$, interchange coefficients, ($l$ = 1, Gibbs energy; $l$ = 2, enthalpy; $l$ = 3, heat capacity), for (b,h) contacts[a] in naphthalene, or biphenyl + 1-alkanol mixtures.

| 1-alkanol | $C^{\mathrm{DIS}}_{\mathrm{bh},1}$ | $C^{\mathrm{DIS}}_{\mathrm{bh},2}$ | $C^{\mathrm{DIS}}_{\mathrm{bh},3}$ | $C^{\mathrm{QUAC}}_{\mathrm{bh},1}$ | $C^{\mathrm{QUAC}}_{\mathrm{bh},2}$ | $C^{\mathrm{QUAC}}_{\mathrm{bh},3}$ |
|---|---|---|---|---|---|---|
| naphthalene | | | | | | |
| 1-tetradecanol | 22.5 | 0.25[b] | −39[b] | 8.9 | 16.7 | 21.2 |
| 1-hexadecanol | 27.5 | 1.4[b] | −39[b] | 8.9 | 16.7 | 21.2 |
| biphenyl | | | | | | |
| 1-tetradecanol | 24 | 0.25[b] | −39[b] | 8.9 | 16.7 | 21.2 |
| 1-hexadecanol | 29 | 1.4[b] | −39[b] | 8.9 | 16.7 | 21.2 |

[a]Type b, aromatic in polycyclic aromatic molecules; type h, OH in 1-alkanols; [b]estimated values

**TABLE 7**

Differences between experimental and theoretical SLE results, $\Delta(T_{\mathrm{SLE}})$ (equation 14) and $\sigma_r(T_{\mathrm{SLE}})$ (equation 15) for PAH + 1-alkanol mixtures obtained from the application of different models

| deviation | $N$ | IDEAL | EQ. (6) | EQ. (12) | DQ[a] | UNIFAC[b] |
|---|---|---|---|---|---|---|
| | | | naphthalene + methanol[c] | | | |
| $\Delta(T_{\mathrm{SLE}})$ /K | 15 | 45 | 68 | 49.4 | 1.5 | 7.6 |
| $\sigma_r(T_{\mathrm{SLE}})$ | 15 | 0.17 | 0.22 | 0.19 | 0.005 | 0.036 |
| | | | naphthalene + 1-butanol[d] | | | |
| $\Delta(T_{\mathrm{SLE}})$ /K | 12 | 46 | 47 | 40 | 0.6 | 8.8 |
| $\sigma_r(T_{\mathrm{SLE}})$ | 12 | 0.15 | 0.16 | 0.10 | 0.002 | 0.032 |
| | | | naphthalene + 1-hexanol[d] | | | |
| $\Delta(T_{\mathrm{SLE}})$ /K | 11 | 34 | 34 | 30 | 0.6 | 9.6 |
| $\sigma_r(T_{\mathrm{SLE}})$ | 11 | 0.11 | 0.11 | 0.10 | 0.002 | 0.034 |
| | | | naphthalene + 1-octanol[d] | | | |
| $\Delta(T_{\mathrm{SLE}})$ /K | 18 | 27 | 27 | 24 | 0.5 | 9.9 |
| $\sigma_r(T_{\mathrm{SLE}})$ | 18 | 0.089 | 0.091 | 0.079 | 0.002 | 0.035 |
| | | | naphthalene + 1-tetradecanol[e] | | | |
| $\Delta(T_{\mathrm{SLE}})$ /K | 9 | 5.8 | 7.1 | 6.2 | 1.2 | 3.5 |
| $\sigma_r(T_{\mathrm{SLE}})$ | 9 | 0.023 | 0.028 | 0.026 | 0.005 | 0.013 |
| | | | naphthalene + 1-hexadecanol[e] | | | |
| $\Delta(T_{\mathrm{SLE}})$ /K | 11 | 4.2 | 7.7 | 6.5 | 1.8 | 1.8 |
| $\sigma_r(T_{\mathrm{SLE}})$ | 11 | 0.018 | 0.023 | 0.020 | 0.008 | 0.007 |
| | | | biphenyl + 1-tetradecanol[e] | | | |
| $\Delta(T_{\mathrm{SLE}})$ /K | 9 | 7.3 | 8.4 | 7.4 | 0.72 | 1.7 |
| $\sigma_r(T_{\mathrm{SLE}})$ | 9 | 0.030 | 0.035 | 0.031 | 0.003 | 0.007 |
| | | | biphenyl + 1-hexadecanol[e] | | | |
| $\Delta(T_{\mathrm{SLE}})$ /K | 10 | 4.8 | 5.4 | 4.8 | 0.77 | 1.1 |
| $\sigma_r(T_{\mathrm{SLE}})$ | 10 | 0.020 | 0.023 | 0.021 | 0.003 | 0.004 |

[a]results obtained using parameters listed in Table 6 and from [26]; [b]results determined using interaction parameters from the literature [25]; [c][12]; [d][11]; [e]this work

**TABLE 8**

Eutectic points for polycyclic aromatic compound(1) + 1-alkanol(2) mixtures. Comparison of experimental (Exp.) results from Wilson and NRTL equations or from the application of DISQUAC (DQ) and UNIFAC models.

| | Exp. | IDEAL | Wilson[a] | NRTL[a] | DQ[b] | UNIFAC[b] |
|---|---|---|---|---|---|---|
| | | | naphthalene + 1-tetradecanol | | | |
| $x_{1eu}$ | 0.208 | 0.344 | 0.167 | 0.177 | 0.250 | 0.172 |
| $T_{eu}/K$ | 306.8 | 304.0 | 307.2 | 307.1 | 306.3 | 308.2 |
| | | | naphthalene + 1-hexadecanol | | | |
| $x_{1eu}$ | 0.271 | 0.432 | 0.229 | 0.235 | 0.327 | 0.266 |
| $T_{eu}/K$ | 316.5 | 313.2 | 317.8 | 318.2 | 316.1 | 317.9 |
| | | | biphenyl + 1-tetradecanol | | | |
| $x_{1eu}$ | 0.201 | 0.415 | 0.217 | 0.174 | 0.252 | 0.184 |
| $T_{eu}/K$ | 307.2 | 302.2 | 309.8 | 307.4 | 306.6 | 308.5 |
| | | | biphenyl + 1-hexadecanol | | | |
| $x_{1eu}$ | 0.267 | 0.512 | 0.248 | 0.263 | 0.360 | 0.298 |
| $T_{eu}/K$ | 316.8 | 311.0 | 317.7 | 317.3 | 315.8 | 317.6 |

[a]results obtained using parameters listed in Table 5; [b]results obtained using interaction parameters from Table 6; [c]results determined using interaction parameters from the literature [25]